\documentclass{article}

\usepackage[english]{babel}

\usepackage[letterpaper,top=2cm,bottom=2cm,left=3cm,right=3cm,marginparwidth=1.75cm]{geometry}

\usepackage[normalem]{ulem}

\usepackage{comment}

\usepackage{tasks}

\usepackage{graphicx}
\graphicspath{{./figures}}
\usepackage{caption}
\usepackage{subcaption}

\usepackage{amsmath}
\usepackage{amsthm}
\usepackage{amsfonts}
\usepackage{amssymb}
\usepackage{mathtools} 
\usepackage{bm}
\usepackage{bbm} 
\usepackage{nicematrix}

\newtheorem{remark}{Remark}

\newcommand{\e}{\mathrm{e}}
\newcommand{\E}{\mathbb{E}}
\newcommand{\F}{\mathcal{F}}
\renewcommand{\i}{\mathrm{i}}
\newcommand{\M}{\mathcal{M}}
\newcommand{\N}{\mathbb{N}}
\newcommand{\R}{\mathbb{R}}

\newcommand{\dt}{\delta t}

\usepackage{scalerel,stackengine}
\stackMath
\newcommand\reallywidehat[1]{
\savestack{\tmpbox}{\stretchto{
  \scaleto{
    \scalerel*[\widthof{\ensuremath{#1}}]{\kern-.6pt\bigwedge\kern-.6pt}
    {\rule[-\textheight/2]{1ex}{\textheight}}
  }{\textheight}
}{0.5ex}}
\stackon[1pt]{#1}{\tmpbox}
}

\usepackage{tikz}
\usetikzlibrary{decorations.markings}
\usetikzlibrary{arrows.meta} 

\tikzset{>=latex} 
\tikzstyle{node}=[thick,circle,draw=black,minimum size=22,inner sep=0.5,outer sep=0.6]
\tikzstyle{node in}=[node,black,draw=black,fill=white]
\tikzstyle{node hidden}=[node,black,draw=black,fill=white]
\tikzstyle{node out}=[node,black,draw=black,fill=white]
\tikzstyle{connect}=[thick,black] 
\tikzstyle{connect arrow}=[-{Latex[length=4,width=3.5]},thick,black,shorten <=0.5,shorten >=1]
\tikzset{ 
  node 1/.style={node in},
  node 2/.style={node hidden},
  node 3/.style={node out},
}
\def\nstyle{int(\lay<\Nnodlen?min(2,\lay):3)} 

\usepackage{pgfplots}
\pgfplotsset{compat=1.17}

\usepackage{float}
\usepackage{algorithm}
\usepackage{algpseudocode}

\usepackage{booktabs} 
\usepackage{multirow} 

\let\savedD\d
\RequirePackage[colorlinks,citecolor=blue!70!green,urlcolor=blue, linkcolor=blue!70!green]{hyperref}
\let\d\savedD
\renewcommand{\d}{\mathrm{d}}

\usepackage[round]{natbib}

\title{Joint deep calibration of the 4-factor PDV model}

\newcommand{\authorblock}[1]{\begin{tabular}{@{}c@{}}#1\end{tabular}}
\author{\begin{tabular}{c@{\qquad}c}
  \authorblock{Fabio Baschetti\\ \scriptsize{Department of Economics, University of Verona, Italy} \\ \footnotesize{\texttt{fabio.baschetti@univr.it}}} &
  \authorblock{Giacomo Bormetti \\ \scriptsize{Department of Economics and Management, } \\ 
  \scriptsize{University of Pavia, Italy} \\
  \footnotesize{\texttt{giacomo.bormetti@unipv.it}}} \\ 
  \multicolumn{2}{c}{\authorblock{Pietro Rossi \\ \scriptsize{Department of Statistical Sciences, University of Bologna, Italy} \\ \footnotesize{\texttt{pietro.rossi3@unibo.it}} \\
  \scriptsize{Prometeia S.p.A., Bologna, Italy} \\
  \footnotesize{\texttt{pietro.rossi@prometeia.com}}}}
\end{tabular}}

\date{\today}

\begin{document}

\maketitle

\begin{abstract}
\noindent Joint calibration to SPX and VIX market data is a delicate task that requires sophisticated modeling and incurs significant computational costs. The latter is especially true when pricing of volatility derivatives hinges on nested Monte Carlo simulation. One such example is the 4-factor Markov Path-Dependent Volatility (PDV) model of \cite{GL:23}. Nonetheless, its realism has earned it considerable attention in recent years. \cite{GG:25} marked a relevant contribution by learning the VIX as a random variable, i.e., a measurable function of the model parameters and the Markovian factors. A neural network replaces the inner simulation and makes the joint calibration problem accessible. However, the minimization loop remains slow due to expensive outer simulation. The present paper overcomes this limitation by learning SPX implied volatilities, VIX futures, and VIX call option prices. The pricing functions reduce to simple matrix–vector products that can be evaluated on the fly, shrinking calibration times to just a few seconds.

\end{abstract}


\noindent \textbf{Acknowledgements:} We acknowledge the CINECA award under the ISCRA initiative, for the availability of high-performance computing resources and support.
We warmly thank Guido Gazzani for fruitful discussions, as well as the participants to the Workshop on Numerical Methods
for Finance and Insurance (University of Milan, 2025), the Financial Engineering Workshop (Bayes Business School, City St George's University of London, 2025), the Workshop on Quantitative Finance (University of Palermo, 2025), and the Vienna Congress on Mathematical Finance (WU Wien, 2025) for insightful comments.

\section{Introduction}

Recent years have brought a deeper understanding of the role of path dependence in volatility modeling. Extensive empirical studies (\citet{Zum:09}, \citeyearpar{Zum:10}, \cite{CB:14}, \cite{BDB:17}) demonstrate that realized volatility depends on the historical price path of the index. \cite{GL:23} formalize this dependence by showing that a simple linear form in $R_1$ and $\sqrt{R_2}$ has the highest predictive power when tested against the level (market value) of the VIX. Here, $R_1$ and $R_2$ are weighted averages of past simple/squared daily SPX returns with different time-shifted power-law (TSPL) kernels. Not only is historical volatility path-dependent, but implied volatility is path-dependent too, according to \cite{Guy:14} and \cite{ABJ:23}. This fact is particularly natural when considering classical stochastic volatility models, where a certain degree of path dependence arises implicitly through non-zero spot-volatility correlation, $\rho$. Conversely, PDV functions as the one described above are fully explicit and thus much more transparent. One layer of flexibility is lost when fixing $\rho=-1$ in the PDV regime. Still, joint SPX/VIX calibration of stochastic volatility models seems to point precisely in that direction, with \cite{GM:22} proving that a highly parameterized neural SDE automatically selects degenerate correlations. The same phenomenon occurs when fitting the 2-factor Bergomi model to the at-the-money SPX skews, and VIX implied volatilities (see \cite{Ber:05}). Regarding the inclusion of an additional exogenous stochastic component, we refer, for example, to \cite{Par:23}. Earlier contributions to the concept of path-dependence in volatility modeling also include \cite{Sen:95}, \cite{HR:98}, and \cite{FP:05}. \\

On the physical side, PDV models reproduce many well-known stylized facts of the volatility process, including the leverage effect, volatility clustering, and the Zumbach effect. The aforementioned linear model is 
no exception. However, one needs a semi-parabolic specification for meaningful pricing under the risk-neutral measure. This brings \cite{GL:23} to the following form of the instantaneous volatility 
\begin{equation*}
    \sigma_t = \beta_0 + \beta_1 R_{1,t} + \beta_2 \sqrt{R_{2,t}} + \beta_{1,2} R_{1,t}^2 \mathbbm{1}_{R_{1,t}\geq 0}
\end{equation*}
in the continuous-time version of the model. Because Markovianity is essential for efficient simulation, one replaces TPLS kernels with convex combinations of two exponentials (two for $R_1$ and two for $R_2$), which fact gives rise to the so-called 4-factor Markov PDV model. \\

While \cite{GL:23} provide sufficient evidence of the flexibility of the model for pricing SPX and VIX derivatives, a proper calibration was missing before \cite{GG:25}. The computational burden associated with this task is indeed substantial, thus demanding a clever workaround for efficient valuation of VIX options and futures that circumvents the need for nested Monte Carlo simulation. In particular, $VIX_T^2 = \frac{1}{\Delta} \E \left[ \int_T^{T+\Delta} \sigma_t^2 \d t \big| \F_T \right]$, and the conditional expectation on the right-hand-side can be learned as a measurable function of the model parameters ($\bm{\theta}$) and the Markovian factors $R$. The natural candidate for the job is a deep, feed-forward neural network. The value of the VIX depends on the integration range only through its length $\Delta$, and this allows the factors to be sampled independently from $T$. In this sense, $T$ is just a label in the map $f: (\bm{\theta},R_T) \mapsto VIX_T^2$. The network effectively replaces the inner simulation so that each outer path is immediately associated with one value of the VIX by evaluating the optimal neural approximation $f^{NN}$ at $(\bm{\theta}, R_T)$. The pricing of volatility derivatives consequently reduces to a standard (i.e., one-level) Monte Carlo. It is a huge step forward, but joint calibration remains very slow (12-20 minutes on a GPU). \\

This paper focuses on eliminating the outer simulation, thereby reducing calibration times and making the joint problem accessible for fast valuation of derivatives in a trading desk. \\

The idea is that we learn the pricing maps directly. The joint problem requires two networks: one network learns SPX implied volatilities, and the other one learns VIX option and futures prices. Dealing with the neural approximation of the former is not that different from what we did in \cite{BBR:24}, the tweaks being entirely model-specific. The 4-factor Markov PDV model is, in fact, much richer than those we considered there -- rough Heston (\cite{GJR:18}) and rough Bergomi (\cite{BFG:16}), and it ultimately produces a wider range of possible smiles with very different levels and shapes. We, therefore, need to train the network learning parameter sets of practical interest. The latter map is significantly more challenging, as the joint problem necessitates the accurate approximation of VIX futures for inversion to VIX implied volatilities (derived from call option prices) to be reliable. Consistency between the futures and the call prices also imposes they come from the same network. \\

Since we train the network on model generated data, we can perform both inner and outer simulations entirely offline, meaning that they never enter the calibration loop. We conveniently confine the Monte Carlo computational burden to the generation phase of the training samples. Simple matrix-vector products, whose evaluation is instantaneous, replace the pricing functions, and the joint calibration takes a few seconds at most. \\

The computational cost of the generation phase is crucial to the success of our strategy. Indeed, it is trivially true that we cannot afford to price hundreds of thousands of VIX surfaces by plain nested simulation, as the number of flops scales with the product of the number of outer and inner paths. To overcome this issue, we leverage least squares regression techniques to achieve considerable speed-up. More specifically, we show that polynomials provide a suitable basis for regression, and an $L_2$ penalty ensures the stability of the coefficients. \cite{GG:21} experiment with more complicated regression models from machine learning; we do believe no additional complexity is needed -- nor convenient -- for our purposes. \\

Our approach is very different from \cite{GG:25}, where
\begin{itemize}
    \item maintaining direct access to the VIX while learning it path by path is extremely convenient for pricing exotics on the volatility itself (e.g., barrier, Bermudan, and American options), but it prevents fast calibration to European payoffs;  
    \item on the contrary, we lose direct access to the VIX as a function of the factors, but we eliminate the need for online simulation. Of course, deep pricing also applies to path-dependent payoffs if needed.
\end{itemize}
Nonetheless, one may use both strategies in combination: $f^{NN}(\bm{\theta}, R_T;\omega^\star)$ could replace the ridge regression and provide a universal representation of the VIX to price options and futures throughout the generation. Universality lies in the fact that optimal neural network weights $\omega^\star$ are independent of the model parameters and the specific realization of the factors. Regression coefficients, on the other hand, are conditional on the pair $(\bm{\theta}, R_T)$ and need to be re-estimated at every outer loop.\\

\cite{BBR:24} is our starting point because:
\begin{enumerate}
    \item it learns a pointwise map 
    \begin{equation*}\label{eqn:spx_nn_map}
        \Phi^{\M,SPX}_p(\bm{\theta},T,K) : (\bm{\theta},T,K) \mapsto \sigma^{\M,SPX}_{BS}(\cdot)
    \end{equation*} 
    which one can use to price any desired strike-maturity pair without  interpolation/extrapolation from a pre-specified grid; 
    \item It demonstrates improved performance compared to standard pointwise techniques, such as \cite{BS:18} and \cite{LBGO:19}.
\end{enumerate}
Keeping $(T, K)$ as inputs is even more important for VIX derivatives, as the true pricing function is far more costly than for SPX options. Having a neural network pricer that evaluates on the fly is essential for prompt trading decisions. The pointwise approach improves on the grid-based methods such as those by \cite{HMT:21} and \cite{Rom:22}. Its flexibility marks the difference between deep pricing and deep calibration: One can calibrate with a neural pricer, but pricing with a grid-based calibration network is much more tricky. \\

We experiment with rough volatility models in \cite{BBR:24}. In particular, we exploit fast Fourier inversion via the SINC approach -- see \cite{baschetti2022sinc} -- when generating training samples from the rHeston model. As for rBergomi, generation hinges on slow Monte Carlo integration. We are forced to this latter route with the 4-factor Markov PDV model. Of course, extension to VIX derivatives makes the generation phase much heavier. \\

Joint SPX/VIX calibration is now the subject of extensive research. In what follows, we mention a few important contributions without any intention of being exhaustive. \cite{PPR:18} improve the flexibility of the 2-factor Heston with jumps and co-jumps by adding a deterministic shift. \cite{RZ:21} calibrate the quadratic rough Heston model of \cite{GJR:20} via a double grid in $(T_{SPX},K_{SPX})$ and $(T_{VIX},K_{VIX})$. \cite{Rom:22} elaborates on multi-factor specifications, possibly rough, with hyperbolic transformations. \cite{BPS:24} adds Hawkes-type jumps to the rough Heston model for increased flexibility. \cite{AIL:25} obtain impressive results using one-factor Gaussian polynomial volatility models. \cite{GLOW:22} devise a non-linear optimal transport approach to the joint calibration problem. \cite{GM:22} employ highly parameterized neural SDEs. \cite{CGS:22} and \cite{CGMS:24} deploy the technology of signatures. \cite{Guy:24} and \cite{BG:24} build non-parametric models by means of minimum entropy techniques.\\


We organize the paper as follows. Section \ref{sec:model} introduces the model, discusses its origins, merits, and limitations, and suggests a simple discretization scheme for simulation over a discrete fine time grid.
In Section \ref{sec:vix}, we recall the definition of the VIX and standard derivatives written on it. We detail robust algorithms for pricing options and futures, tailor them to the specific case of the 4-factor Markov PDV model, and demonstrate how least squares regression can effectively improve upon nested Monte Carlo simulation. Section \ref{sec:nn} guides the reader through the core of the manuscript. We present our neural approximation for VIX derivative prices, from the generation of samples and learning procedure to the out-of-sample performance of the network. Section \ref{sec:calib} introduces the loss function and summarizes an extensive calibration exercise (both on the SPX surface and the joint problem). Section \ref{sec:concl} concludes.

\section{The 4-factor Markov PDV model} \label{sec:model}

The pioneering work by \cite{GL:23} attempts to model the explicit dependence of implied and realized volatility of equity indexes on past log-returns. The idea is to specify a simple yet effective functional form for reconstructing the volatility behavior. Minor modifications follow for meaningful pricing in continuous time. \\

In this section, we retrace the complete history of the model, discuss the key ideas, and propose a straightforward simulation scheme.

\subsection{Motivation}

Following the original paper, we first introduce the two fundamental building blocks of any reasonable volatility function, namely:
\begin{enumerate}
    \item a trend feature
    \begin{equation*}
        R_{1,t} = \sum_{t_i \leq t} K_1(t-t_i)r_{t_i}
    \end{equation*}
    that describes recent trends in asset prices to capture the leverage effect, i.e., the phenomenon where volatility tends to rise when asset prices fall and, more generally, the Zumbach effect.
    \item An activity feature
    \begin{equation*}
        R_{2,t} = \sum_{t_i \leq t} K_2(t-t_i)r_{t_i}^2
    \end{equation*}
    capturing recent activity in the asset squared returns to describe volatility clustering, i.e., the phenomenon where periods of large (small) volatility tend to follow periods of large (small) volatility.
\end{enumerate}

The variable
\begin{align*}
    r_{t_i} = \frac{S_{t_i}-S_{t_{i-1}}}{S_{t_{i-1}}}
\end{align*}
denotes the daily return, and $K_n: \R_{\geq 0} \mapsto \R_{\geq 0}, \, n = \{1,2\}$ are convolution kernels. \\

A multivariate lasso regression selects a time-shifted power law
\begin{equation*}
    K(\tau) = \frac{\alpha - 1}{\delta^{1-\alpha}}(\tau+\delta)^{-\alpha}
\end{equation*}
with $\tau \geq 0, \alpha >1, \delta>0$, for both $n=1$ and $n=2$. Such specification effectively mixes short and long memory as: i) recent returns receive much higher weights than older returns, but ii) the weights decrease very slowly for large $\tau$ (way more slowly than an exponential decay). Crucially, then, the time shift $\delta$ is strictly positive and guarantees that the kernel does not explode in the limit $\tau \to 0$, which fact marks an important difference with respect to rough volatility. \\

The authors show that a very simple model of the form 
\begin{equation}\label{mdl:emp_dtim_pdv}
    \sigma_t = \beta_0 + \beta_1 R_{1,t} + \beta_2 \sqrt{R_{2,t}}    
\end{equation}
explains up to 90\% of the variability in the VIX and about 60\% of the variability in future realized volatility (RV). Of course, $\beta_0>0$. Then $\beta_1<0$ is needed to produce a leverage effect, and $\beta_2 \in (0,1)$ generates volatility clustering\footnote{$\beta_2 < 1$ is a technical requirement for stability. Any $\beta_2>0$ would work given the desired clustering phenomenon.}. \\

The continuous-time limit of model (\ref{mdl:emp_dtim_pdv}) can be readily written as
\begin{align*}
    \frac{\d S_t}{S_t} & = \sigma_t \d W_t \\
    \sigma_t & = \sigma(R_{1,t},R_{2,t}) \\
    \sigma(R_1,R_2) & = \beta_0 + \beta_1 R_1 + \beta_2 \sqrt{R_2} \\
    R_{1,t} & = \int_{-\infty}^t K_1(t-u) \frac{\d S_u}{S_u} =  \int_{-\infty}^t K_1(t-u) \sigma_u \d W_u \\
    R_{2,t} & = \int_{-\infty}^t K_2(t-u) \left( \frac{\d S_u}{S_u} \right)^2 \\ & = \int_{-\infty}^t K_2(t-u) \sigma_u^2 \d u
\end{align*}
but one immediately recognizes a couple of important aspects:
\begin{itemize}
    \item the model is non-Markovian, in general. However, one can approximate a TSPL  by a convex combination of two exponential functions with different decays
    \begin{equation*}
        K_{\theta,\lambda_0,\lambda_1}(\tau) : \tau \mapsto (1-\theta) \lambda_0 \e^{-\lambda_0\tau} + \theta \lambda_1 \e^{-\lambda_1\tau} \hspace{5mm} \text{where} \hspace{2mm} \theta \in [0,1], \, \lambda_0 > \lambda_1 > 0.
    \end{equation*}
    
    The authors further decompose the factors as 
    \begin{align*}
        & R_{1,t} = (1-\theta_1) R_{1,0,t} + \theta_1 R_{1,1,t} \\
        & R_{2,t} = (1-\theta_2) R_{2,0,t} + \theta_2 R_{2,1,t}
    \end{align*}
    and the individual pieces 
    \begin{align*}
        & R_{1,j,t} = 
        \e^{-\lambda_{1,j}t}R_{1,j,0} + \int_0^t \lambda_{1,j} \e^{-\lambda_{1,j}(t-u)}\sigma_u \d W_u & j \in \{0,1\} \\
        & R_{2,j,t} = 
        \e^{-\lambda_{2,j}t}R_{2,j,0} + \int_0^t \lambda_{2,j} \e^{-\lambda_{2,j}(t-u)}\sigma_u^2 \d u & j \in \{0,1\} 
    \end{align*}
    explicitly depend on their initial values. 
    \item $R_2$ boils down to the solution of an ordinary differential equation. When going to continuous time,  one obtains
    \begin{align*}
        r_{t_i} = \frac{S_{t_i}-S_{t_{i-1}}}{S_{t_{i-1}}} \longrightarrow \frac{\d S_t}{S_t} = \sigma_t \d W_t \quad \implies \quad r_{t_i}^2 \longrightarrow \left(\frac{\d S_t}{S_t}\right)^2 = \sigma_t^2 \d t = \sigma_t^2 \langle \d W_t \rangle\,.
    \end{align*}
    The square of the return is much more informative than the quadratic variation of the innovation. Since $r_{t_i}^2$ marks a big difference in comparison to $\Delta t_i = \E[r_{t_i}^2]$, one must regretfully conclude that the continuous time version is much less informative about the volatility dynamics than the initial discrete time specification. \\
    
    As a result, the marginal distribution of the spot price suffers from a severe lack of probability mass in the left tail, i.e., SPX smiles are too flat compared to the market. A possible solution for this is artificially pumping volatility by adding a parabolic component active when a positive price trend appears
    \begin{equation*}
        \sigma_t = \beta_0 + \beta_1 R_{1,t} + \beta_2 \sqrt{R_{2,t}} + \beta_{1,2} R_{1,t}^2 \mathbbm{1}_{R_{1,t}\geq 0}.
    \end{equation*}
    While this specification adds one parameter $\beta_{1,2} \geq 0$, it works effectively only in a small region around the ATM; then, the smile's right wing grows too high. \\

    We provide a graphical description of the effect of $\beta_{1,2}$ on both SPX and VIX smiles in Figure \ref{fig:effect_beta12}. 

    \begin{figure}[!ht]
    \centering
    \begin{subfigure}[b]{0.49\textwidth}
        \centering
        \includegraphics[width=\linewidth]{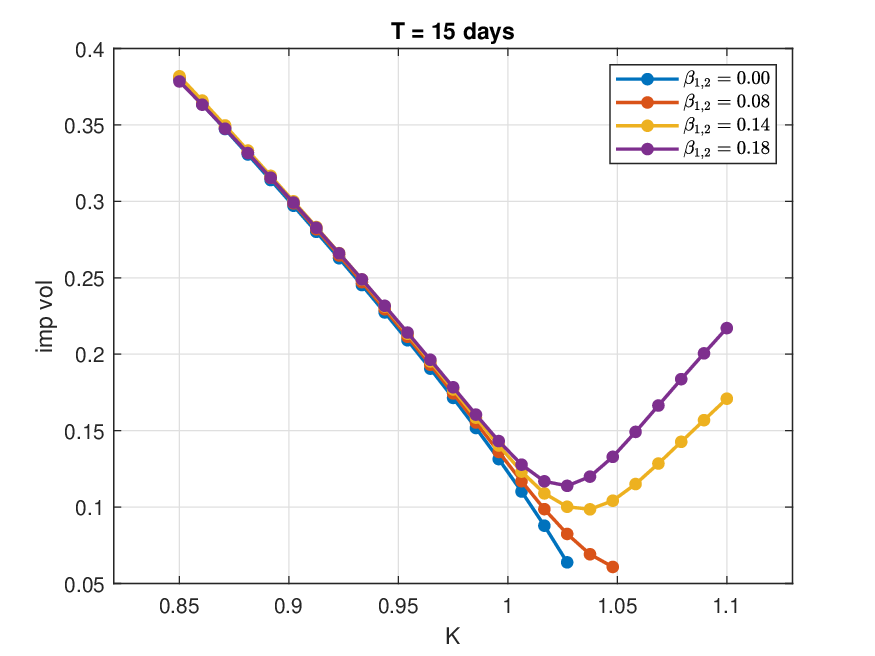}
    \end{subfigure}
    \begin{subfigure}[b]{0.49\textwidth}
        \centering
        \includegraphics[width=\linewidth]{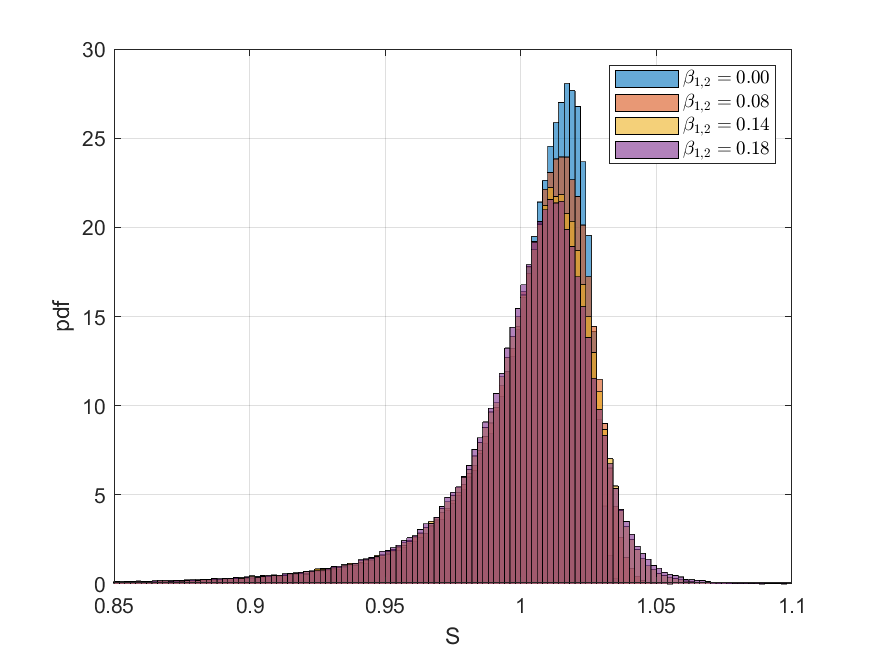}
    \end{subfigure}
    \begin{subfigure}[b]{0.49\textwidth}
        \centering
        \includegraphics[width=\linewidth]{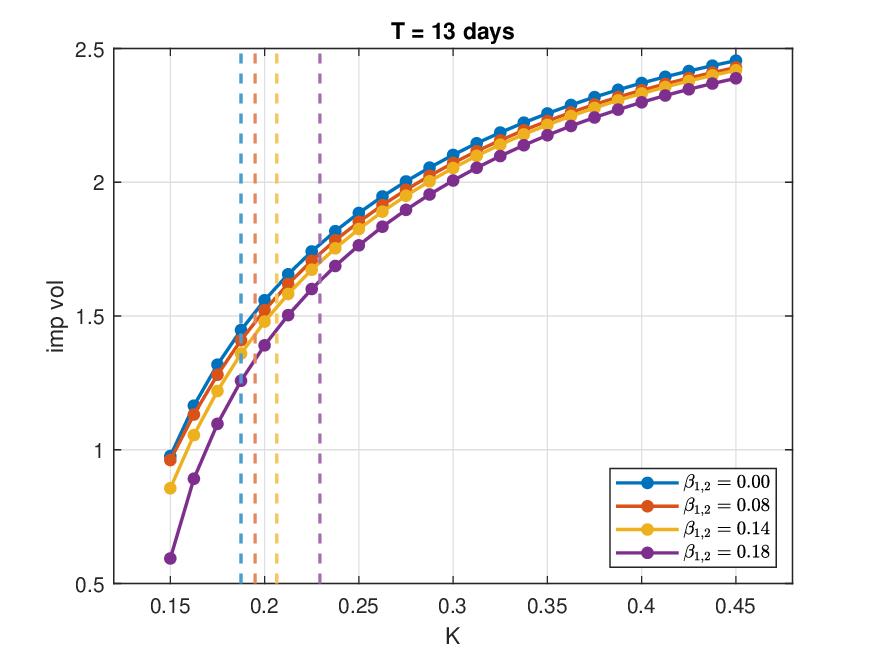}
    \end{subfigure}
        \caption[4-factor Markov PDV model smiles and terminal-price histograms as a function of $\beta_{1,2}$]{4-factor Markov PDV model smiles and terminal-price histograms as a function of $\beta_{1,2}$. Call options on the SPX are valued by plain Monte Carlo with $N=2^{18}$ paths; VIX derivatives require nested simulation: $\mathrm{N^\text{sim}_\text{out}}=2^{18}$, $\mathrm{N^\text{sim}_\text{inn}}=2^{13}$ (the notation will be clarified in the main text). In both cases, we use six time-steps per day. Model parameters: $\beta = (0.026 , -0.165 , 0.680 , \beta_{1,2})$,  $\lambda_1 = (48.11 , 39.52)$, $\theta_1 = 0.385$, $\lambda_2 = (9.59 , 3.31)$,  $\theta_2 = 0.910$. Factor initial values: $R_{1,0} = (0.0636, 0.0874)$, $R_{2,0} = (0.0159, 0.0266)$.}
    \label{fig:effect_beta12}
    \end{figure}

    \cite{GG:25} elaborate extensively on this aspect and we refer the reader to it for any additional detail. \\

    Alternatively, one can account for the lack of probability on the tails by multiplying the PDV function by an exogenous stochastic component 
    \begin{equation*}
        \sigma_t = \e^{X_t} \left( \beta_0 + \beta_1 R_{1,t} + \beta_2 \sqrt{R_{2,t}} \right),
    \end{equation*}
    where $X_t$ may follow an Ornstein-Uhlenbeck process
    \begin{equation*}
        X_t = -\kappa X_t \d t + \nu \d Z_t, \qquad Z_t \perp W_t.
    \end{equation*}
    The last solution goes in the direction of the path-dependent stochastic volatility (PDSV) models and it is actually much more realistic than the additive  parabolic correction to the function $\sigma(R_1,R_2)$. Of course, modeling an additional multiplicative process $X_t$ ends up increasing the dimensionality of the calibration problem. 
\end{itemize}

We stick to the parabolic Markov model, which we now describe in full generality.

\subsection{The model}

The 4-factor Markov PDV model reads as follows
\begin{align}
    \frac{\d S_t}{S_t} & = \sigma_t \d W_t \nonumber \\
    \sigma_t & = \sigma(R_{1,t},R_{2,t}) \nonumber \\
    \sigma(R_1,R_2) & = \beta_0 + \beta_1 R_1 + \beta_2 \sqrt{R_2} + \beta_{1,2} R_1^2 \mathbbm{1}_{R_1\geq 0} \label{eqn:pdv_para_fun} \\
    R_{1,t} & = (1-\theta_1) R_{1,0,t} + \theta_1 R_{1,1,t} \label{eqn:R1_cmp}\\
    R_{2,t} & = (1-\theta_2) R_{2,0,t} + \theta_2 R_{2,1,t} \label{eqn:R2_cmp} \\
    \d R_{1,j,t} & = \lambda_{1,j} \left( \sigma(R_{1,t},R_{2,t})\d W_t - R_{1,j,t}\d t \right) \hspace{5mm} j \in \{0,1\} \nonumber \\ 
    \d R_{2,j,t} & = \lambda_{2,j} \left( \sigma(R_{1,t},R_{2,t})^2 - R_{2,j,t} \right) \d t \hspace{9.80mm} j \in \{0,1\} \nonumber
\end{align}
where 
\begin{itemize}
    \item $\beta_0>0$ is the baseline volatility, $\beta_1<0$ and $\beta_2 \in (0,1)$ the coefficients of the empirical model. $\beta_{1,2} \geq 0$ is the (additional) parabolic term
    \item $\lambda_{1,0}>0$ captures the dependence of $R_1$ on recent returns; the higher $\lambda_{1,0}$, the larger the weights of recent returns 
    \item $\lambda_{1,1}<\lambda_{1,0}$ captures the dependence of $R_1$ on older returns; the smaller $\lambda_{1,1}$, the larger the weights of older returns
    \item $\lambda_{2,0}>0$ captures the dependence of $R_2$ on recent squared returns; the higher $\lambda_{2,0}$, the larger the weights of recent squared returns 
    \item $\lambda_{2,1}<\lambda_{2,0}$ captures the dependence of $R_2$ on older squared returns; the smaller $\lambda_{2,1}$, the larger the weights of older squared returns
    \item $\theta_1$ and $\theta_2$ mix the dependence on recent and older returns --- and squared returns.
\end{itemize}
The choice of the two-exponential kernel makes the quadruplet $(R_{1,0,t},R_{1,1,t},R_{2,0,t},R_{2,1,t})$ Markovian. Hence, the name of the model. \\

Most importantly, Markovianity facilitates and speeds up the model simulation.

\subsection{Simulation}

Simulation of the 4-factor Markov PDV model is extremely easy. As we already noticed, the $R_2$ components follow a simple ODE in continuous time. Then, if we assume volatility $\sigma_t$ to be constant over each discretization interval of length $\dt$, we can write
\begin{align*}
    R_{2,j,t+\dt} = \sigma_t^2 - \e^{-\lambda_{2,j}\dt} (\sigma_t^2 - R_{2,j,t})  
\end{align*}
for $j \in \{0,1 \}$. By the same hypothesis -- and a direct application of It\^o's product rule for $\e^{\lambda_{1,j,t}t}R_{1,j,t}$ -- one immediately recovers 
\begin{align*}
    R_{1,j,t+\dt} = \e^{-\lambda_{1,j}\dt} \left(R_{1,j,t} + \lambda_{1,j} \sigma_t \sqrt{\dt} \epsilon \right)
\end{align*}
for $j \in \{0,1 \}$. Needless to say that $\epsilon \sim \mathcal{N}(0,1)$. \\
\begin{algorithm}
\caption{Simulate\_onestep$(\cdot,N)$}
\begin{algorithmic}
\State \% $N$: number of trajectories
\State
\State $\delta W_t = \mathrm{randn}(N,1) \times \sqrt{\dt}$
\State $S_{t+\dt} = S_t \times \mathrm{exp}\left( -\frac{1}{2}\sigma_t^2 \dt + \sigma_t \delta W_t \right)$ 
\State $R_{1,0,t+\dt} = \mathrm{exp}(-\lambda_{1,0} \dt) \times \left( R_{1,0,t} + \lambda_{1,0} \sigma_t \delta W_t \right)$
\State $R_{1,1,t+\dt} = \mathrm{exp}(-\lambda_{1,1} \dt) \times \left( R_{1,1,t} + \lambda_{1,1} \sigma_t \delta W_t \right)$
\State $R_{2,0,t+\dt} = \mathrm{exp}(-\lambda_{2,0} \dt) \times (R_{2,0,t} + \lambda_{2,0} \sigma_t^2 \dt)$
\State $R_{2,1,t+\dt} = \mathrm{exp}(-\lambda_{2,1} \dt) \times (R_{2,1,t} + \lambda_{2,1} \sigma_t^2 \dt)$
\State $[R_{1,t+\dt},R_{2,t+\dt}] = \mathrm{r\_cmp}([R_{1,0,t+\dt},R_{1,1,t+\dt}],[R_{2,0,t+\dt},R_{2,1,t+\dt}])$
\State $\sigma_{t+\dt} = \min(\mathrm{f\_pdv}(R_{1,t+\dt},R_{2,t+\dt}),1.5)$
\end{algorithmic}\label{algo:pdv_onestep}
\end{algorithm}
\\
The algorithm \ref{algo:pdv_onestep} describes the one-step updating rule. Here the function $\textit{randn}$ generates samples from a standard normal random variable, $\textit{r\_cmp}$ returns the current value of the factors according to Equations \eqref{eqn:R1_cmp}-\eqref{eqn:R2_cmp}, and $\textit{f\_pdv}$ updates the volatility based on the PDV function \eqref{eqn:pdv_para_fun}. Please note that, following market practice, we cap the volatility at 1.5. The upper bound is necessary to avoid explosions, but it obviously affects the dynamics of $R_1$ and $R_2$. The value of the cap is quite arbitrary, and we follow \cite{GL:23} in our choice. 

\section{The Volatility Index (VIX)} \label{sec:vix}

The CBOE Volatility Index (VIX) measures the market's expectation of 30-day volatility, as implied by the SPX, through the aggregation of weighted prices of put and call options. The market definition reads as follows
\begin{equation}\label{eqn:vix_mkt}
    VIX_T^2 = \mathrm{Price}_T \left[ -\frac{2}{\Delta} \log \left( \frac{S_{T+\Delta}}{F_T^{T+\Delta}} \right) \right]
\end{equation}
where $F_t^u$ denotes the time-$t$ price of the SPX future expiring at $u \geq t$, and $\Delta = \frac{30}{365}$ is a 30-day horizon. \\

The model counterpart of \eqref{eqn:vix_mkt} is given by
\begin{equation}\label{eqn:vix_mdl}
    VIX_T^2 = \E \left[ - \frac{2}{\Delta} \log \left( \frac{S_{T+\Delta}}{F_T^{T+\Delta}} \right) \Bigg| \F_T \right].
\end{equation}

If there are no jumps in the underlying $S$, then Equation \eqref{eqn:vix_mdl} reduces to
\begin{align}
    VIX_T^2 & = \E \left[ \frac{1}{\Delta} \int_T^{T+\Delta} \sigma_u^2 \, \d u \;  \Bigg| \F_T \right] \label{eqn:vix_Snj} \\
    & = \frac{1}{\Delta} \int_T^{T+\Delta} \E[\sigma_u^2 | \F_T] \, \d u = \frac{1}{\Delta} \int_T^{T+\Delta} \xi_T^u \, \d u \nonumber 
\end{align}
via a straightforward application of It\^o's formula. The normalization by $\Delta$ makes the VIX an average of expected future variance on the SPX. $\xi_t^u = \E[\sigma_u^2 | \F_t]$ is the forward variance curve. \\

Financial markets actively trade VIX options and futures as tools for market participants to manage volatility directly. VIX futures reflect the market's estimate of the volatility index at various expiration dates
\begin{equation*}
    F_{VIX}(t,T) = \E [VIX_T|\mathcal{F}_t].
\end{equation*}
VIX options 
\begin{equation*} 
    C_{VIX}(t,T,K) = \e^{-r(T-t)} \E [(VIX_T - K)^+|\mathcal{F}_t] 
\end{equation*}
enable financial institutions to hedge portfolio volatility risk and trade based on their view of future volatility movements.

\subsection{Pricing VIX derivatives}

Pricing of VIX derivatives is a computationally burdensome operation in the absence of convenient mathematical structures, such as affine and polynomial specifications. In the former case, the availability of generalized Fourier transforms for futures and call option payoffs enables semi closed-form evaluation by the Lewis integral (\citet{Lew:00}, \citeyearpar{Lew:01}). As for polynomial models, the moment formula allows for the computation of conditional mixed moments of the process components in terms of a matrix exponential (see \cite{CKT:12}, \cite{FL:16}). The latter can be readily used for $VIX_T^2$ whenever $\sigma^2$ is a known polynomial function of the components' terminal value, which fact requires preliminary simulation up to time $T$ by standard Monte Carlo. \\

As soon as the polynomial property does not hold -- as is usually the case with most stochastic volatility models -- pricing VIX derivatives may become computationally expensive. In fact they should be treated by nested Monte Carlo, in full generality. The latter requires the three steps described below: \\

\begin{enumerate}
    \item \underline{Outer simulation}: draw $\mathrm{N^\text{sim}_\text{out}}$ independent paths of the state variables $\boldsymbol{M}_t$ up to maturity $T$.   
        \begin{algorithm}
        \caption{Outer simulation}
        \begin{algorithmic}
            \State \% \ $\dt$: time discretization step
            \State \% \ $N^\text{steps}_\text{out} = \frac{T}{\dt}$
            \State \% \ $N^\text{sim}_\text{out}$: number of outer paths
            \State 
            \For{$k=1:\mathrm{N^\text{steps}_\text{out}}$}
            \State $M_{k \times \dt} = \mathrm{simulate\_onestep}(M_{(k-1) \times \dt}, N=\mathrm{N^\text{sim}_\text{out}})$ 
            \EndFor   
        \end{algorithmic}\label{algo:vix_outer}
        \end{algorithm} \\
        For the 4-factor Markov PDV model, we have $\boldsymbol{M}_t = (S_t,R_{1,0,t},R_{1,1,t},R_{2,0,t},R_{2,1,t},\sigma_t)$. 
    \item \underline{Inner simulation}: build $\boldsymbol{M}_T^{(n)}$-conditional discrete volatility trajectories 
    \[
    \begin{array}{ll}
        1 & \sigma_{T}^{(n)}, \sigma_{T+\delta t}^{(n)(1)}, \dots, \sigma_{T+ \Delta}^{(n)(1)} \\
        2 & \sigma_{T}^{(n)}, \sigma_{T+\delta t}^{(n)(2)}, \dots, \sigma_{T+ \Delta}^{(n)(2)} \\
        \vdots & \vdots \\
        \mathrm{N^\text{sim}_\text{inn}} & \sigma_{T}^{(n)}, \sigma_{T+\delta t}^{(n)(\mathrm{N^\text{sim}_\text{inn}})}, \dots, \sigma_{T+ \Delta}^{(n)(\mathrm{N^\text{sim}_\text{inn}})}
    \end{array}
    \]
    over the interval $[T, T+\Delta]$, approximate 
    \begin{equation*}
        \frac{1}{\Delta} \int_T^{T+\Delta} \sigma_u^{2(n)(j)} \d u, \qquad j=1,\dots,\mathrm{N^\text{sim}_\text{inn}}
    \end{equation*}
    with a simple rectangular rule
    \begin{align*}
        \frac{1}{\mathrm{N^\text{steps}_\text{inn}}+1} \sum_{k=0}^{\mathrm{N^\text{steps}_\text{inn}}} \sigma^{2(n)(j)}_{T+k\dt}
    \end{align*}
    and average over the inner paths to estimate of $VIX_T^{2(n)}$. Take the square root for $VIX_T^{(n)}$. Repeat for all $n=1,\dots,\mathrm{N^\text{sim}_\text{out}}$. \\

    Algorithm \ref{algo:vix_inner} provides a pseudo-code implementation.
    \begin{algorithm}
    \caption{Inner simulation}
    \begin{algorithmic}
        \State \% \ $N^\text{sim}_\text{inn}$: number of inner paths
        \State \% \ $N^\text{steps}_\text{inn} = \frac{\Delta}{\dt}$
        \State
        \State $VIX^2$ = zeros($N^\text{sim}_\text{out}$,1)
        \For{$n=1:N^\text{sim}_\text{out}$}
            \State $\Sigma$ = [$\sigma_T^{2(n)}$ zeros($N^\text{sim}_\text{inn}$, $N^\text{steps}_\text{inn}$)]
            \For{$k=1:N^\text{steps}_\text{inn}$}
                \State $M^{(n)}_{T+k \times \dt}$ = simulate\_onestep($M^{(n)}_{T+(k-1) \times \dt}$, $N=$ $N^\text{sim}_\text{inn}$)
                \State $\Sigma(:,k+1) = \sigma^{2(n)}_{T+k \times \dt}$
            \EndFor
            \State $I$ = sum($\Sigma$,2) / ($N^\text{steps}_\text{inn}$ + 1)
            \State $VIX^2(n)$ = mean($I$)
        \EndFor
        \State $VIX$ = sqrt($VIX^2$)
    \end{algorithmic}\label{algo:vix_inner}
    \end{algorithm}
    \item \underline{Pricing}: compute the fair value of any derivative contract by means of a sample average. 
\end{enumerate}

The reader immediately understands that the computational complexity grows very fast as the product of the number of outer and inner paths. Nonetheless, nested Monte Carlo is universally recognized as a robust methodology. It becomes very accurate when $N^\text{sim}_\text{out}$ and $N^\text{sim}_\text{inn}$ are sufficiently large. In what follows, our numerical experiments suggest that 
\begin{align*}
    & N^\text{sim}_\text{out} = 2^{18} \\
    & N^\text{sim}_\text{inn} = 2^{13}
\end{align*} 
is a good choice for computing benchmark high-precision derivative prices. Please note that this choice already amounts to drawing nearly two billion trajectories overall. As this is completely impractical for any real-world application, we need solutions to alleviate the computational burden of the pricing problem. \\

\begin{figure}[!ht]
	\centering
	\begin{subfigure}[b]{0.49\textwidth}
		\centering
		\includegraphics[width=\textwidth]{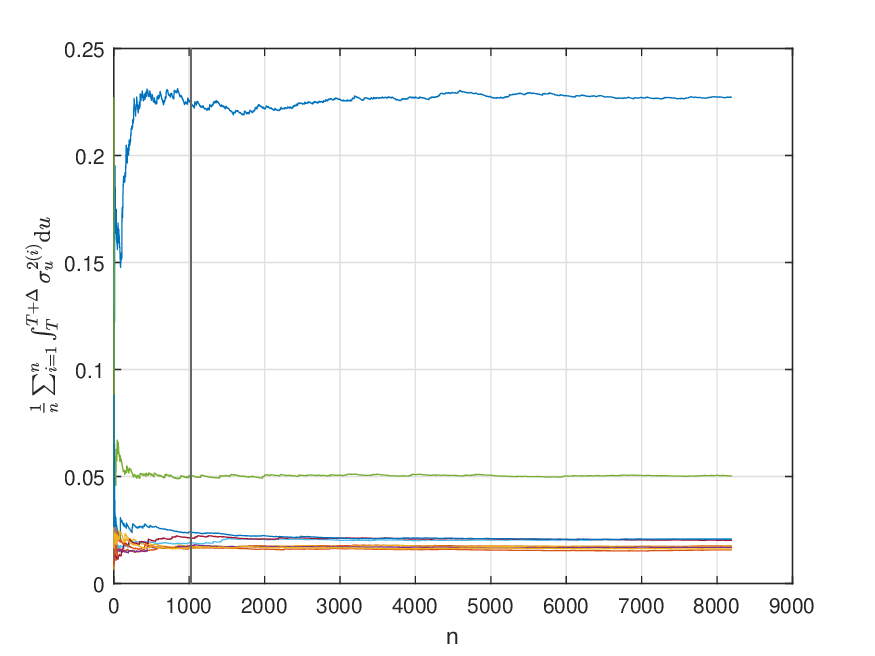}
	\end{subfigure}
	\begin{subfigure}[b]{0.49\textwidth}
		\centering
		\includegraphics[width=\textwidth]{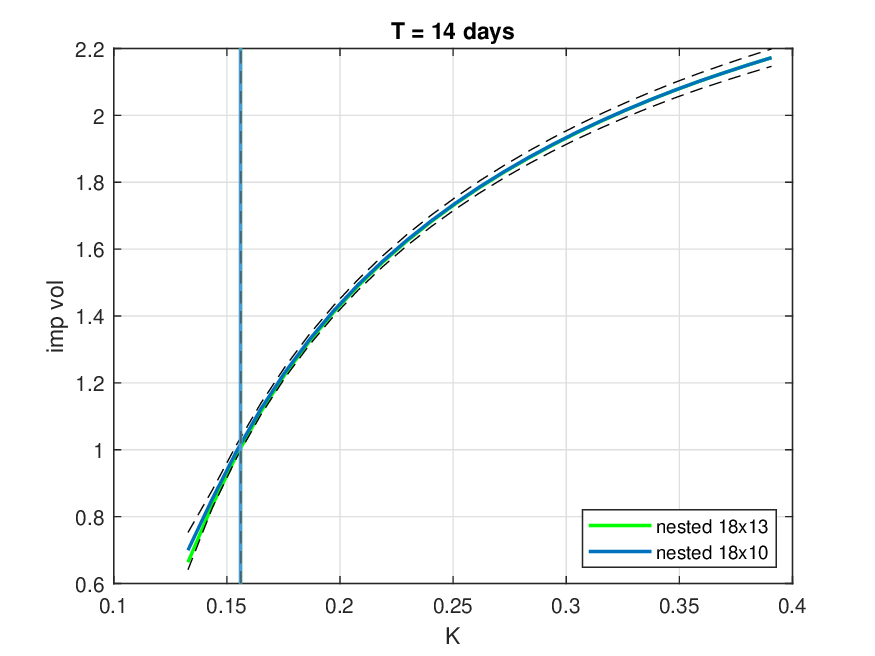}
	\end{subfigure}
         \caption[Convergence of $VIX_T^2$ estimates and its effect on the smile]{Convergence of $VIX_T^2$ estimates and its effect on the smile. LHS: behavior of the Monte Carlo estimate of $VIX_T^2$ as a function of the number of points included in the mean. RHS: comparison between the benchmark smile (green) and the prices obtained when reducing to $\log_2N^\text{sim}_\text{inn}= 10$ (blue). Dashed black lines represent 95\% confidence bands. $\log_2N^\text{sim}_\text{out}=18$ and the number of time steps per day is six. Model parameters: $\beta = (0.031, -0.134, 0.590, 0.108)$,  $\lambda_1 = (61.11, 32.25)$, $\theta_1 = 0.25$, $\lambda_2 = (9.51, 5.23)$,  $\theta_2 = 0.98$. Initial values of the factors: $R_{1,0} = (0.3038, 0.2505)$, $R_{2,0} = (0.0164, 0.0199)$.}
	\label{fig:vix2_inn_enough}
\end{figure}

One key observation is that the inner simulation appears to have quite a low variance, with the left-hand-side of Figure \ref{fig:vix2_inn_enough} showing that about a thousand paths are sufficient for a reliable estimate of $VIX_T^2$. In fact $\frac{1}{n}\sum_{i=1}^n \int_T^{T+\Delta} \sigma_u^{2(i)} \d u$ is pretty stable as a function of $n$ after a few hundred terms have been included. \\

Most importantly, the residual variance in the individual $VIX_T^2$ estimates appears to be averaged out by the outer loop when examining the smile. Benchmark volatilities indeed fit inside the 95\% confidence bands of a reduced run with $N^\text{sim}_\text{inn}=2^{10}$.

While we have seen we can safely reduce the number of inner paths, the true game changer when pricing volatility derivatives is a variant of nested Monte Carlo, which deploys least-squares techniques to learn $VIX_T$ in terms of a polynomial function. \\

The key steps are as follows: 
\begin{enumerate}
    \item[i-] apply the inner loop on a sub-sample of outer paths of dimension $N^\text{sub} \ll N^\text{sim}_\text{out}$ to produce a column vector $\boldsymbol{VIX}_T^\text{sub}$ of dimension $N^\text{sub}$;
    \item[ii-] specify a basis $\boldsymbol{X}$ -- of dimension $\mathrm{dim}(\boldsymbol{X})$ -- of monomials of degree up to $d$ and evaluate it on the same sub-sample of state variables $\boldsymbol{M}_T^\text{sub}$ as previously used for $\boldsymbol{VIX}_T^\text{sub}$. Organize the output in a matrix $X^\text{sub}\in \mathbb{R}^{N^\text{sub}\times \mathrm{dim}(\boldsymbol{X})}$;  
    \item[iii-] solve the ridge regression problem
    \begin{equation*}
        \operatorname*{argmin}_{\boldsymbol{b} \in \R^{\mathrm{dim}(\boldsymbol{X})}} \|  X^\text{sub} \boldsymbol{b} - \boldsymbol{VIX}_T^\text{sub} \|_2^2 + c \| \boldsymbol{b} \|_2^2
    \end{equation*}
    where the $L_2$ penalizer $c>0$ controls the magnitude of the estimated coefficients to prevent overfitting. 
    The closed-form solution reads as
    \begin{equation*}
        \boldsymbol{b}^\star = (X^{\text{sub}\top} X^\text{sub} + c I)^{-1} X^{\text{sub}\top} \boldsymbol{VIX}_T^\text{sub}\,.
    \end{equation*}
    Several algorithms exist to compute the solution in a numerically stable manner. In particular, the QR decomposition is often the preferred choice due to accuracy considerations.  
    \item[iv-] Evaluate the basis $\boldsymbol{X}$ on the full sample of outer simulations and arrange the output in a matrix $X^\text{full}\in \mathbb{R}^{N^\text{sim}_\text{out}\times \mathrm{dim}(\boldsymbol{X})}$; compute the VIX for the whole sample
    \begin{equation*}
        \boldsymbol{VIX}_T^\text{full} = X^\text{full} \boldsymbol{b}^\star\,.
    \end{equation*}
    Additionally, one can only use sample points that did not enter the estimation process. The sake of robustness justifies the additional negligible cost of producing extra outer paths. 
\end{enumerate}

\begin{figure}[!ht]
	\centering
	\begin{subfigure}[b]{0.49\textwidth}
		\centering
		\includegraphics[width=\textwidth]{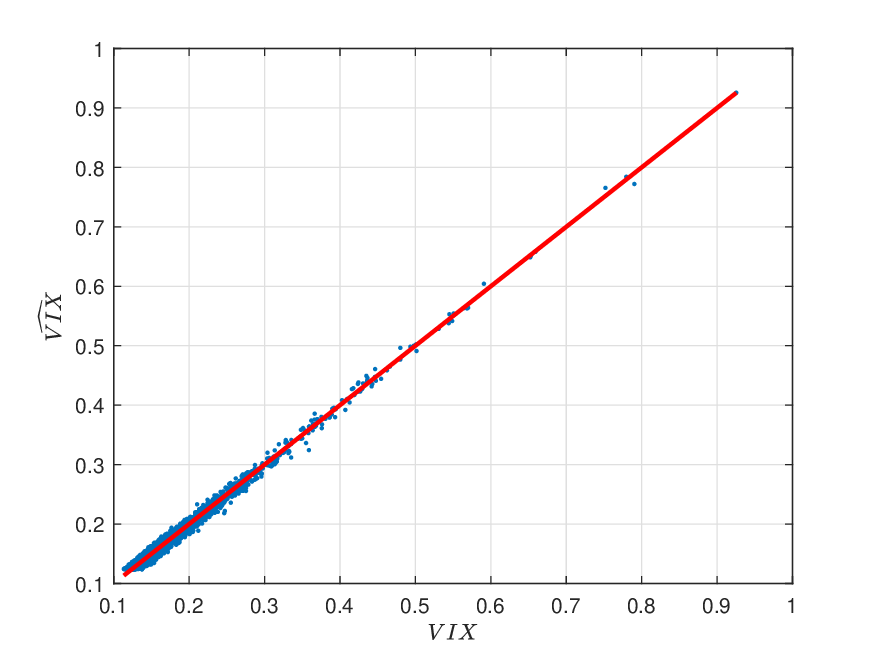}
	\end{subfigure}
	\begin{subfigure}[b]{0.49\textwidth}
		\centering
		\includegraphics[width=\textwidth]{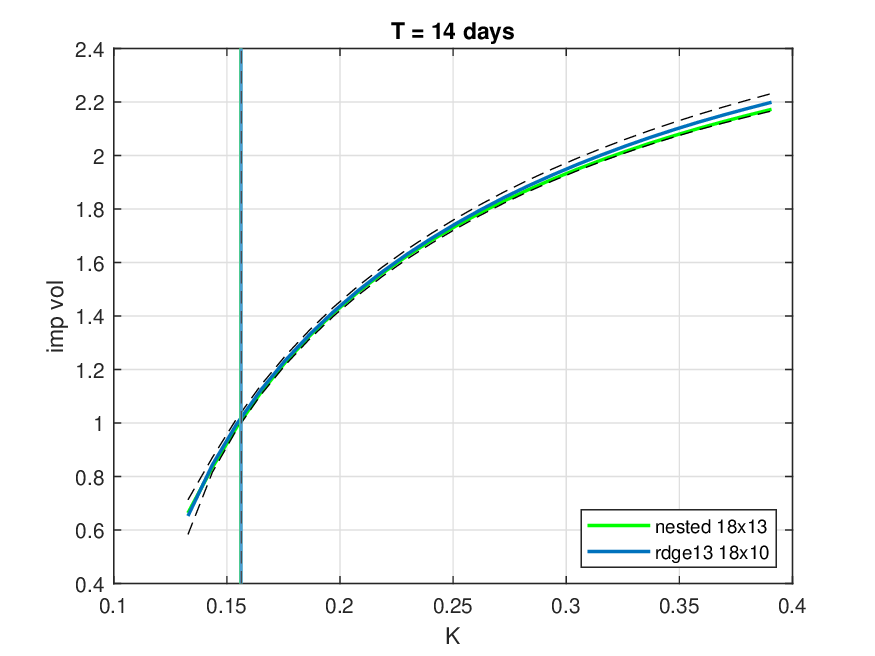}
	\end{subfigure}
         \caption[Quality of the ridge regression for $VIX_T$ and its effect on the smile]{Quality of the ridge regression for $VIX_T$ and its effect on the smile. LHS: $VIX_T$ as computed by inner simulation \textit{vs} its polynomial approximation $\widehat{VIX}_T$. RHS: comparison between the benchmark smile (green) and the prices obtained by Least Squares Monte Carlo with $\log_2 N^{\text{sub}}=13$ and  $\log_2 N^\text{sim}_\text{inn}=10$ (blue). Black lines correspond to 95\% confidence bands. $\log_2N^\text{sim}_\text{out}=18$ and the number of time steps per day is six. Model parameters: $\beta = (0.031, -0.134, 0.590, 0.108)$,  $\lambda_1 = (61.11, 32.25)$, $\theta_1 = 0.25$, $\lambda_2 = (9.51, 5.23)$,  $\theta_2 = 0.98$. Initial values of the factors: $R_{1,0} = (0.3038, 0.2505)$, $R_{2,0} = (0.0164, 0.0199)$.}
	\label{fig:bmark_vs_ols13}
\end{figure}

We demonstrate the quality of the polynomial fit to $VIX_T$ with the scatter plot in Figure \ref{fig:bmark_vs_ols13}. On the other hand, the extension is sufficient for the confidence bands of the Least Squares Monte Carlo run to encompass the benchmark volatilities. 
Here, we estimate the regression coefficients using $N^\text{sub} = 2^{13}$ sample points. The number of inner paths to compute reduces from $2^{31}$ to $2^{13} \times 2^{10} = 2^{23}$  with a gain factor of order 256. This speed-up ensures that we can afford the cost associated with generating a rich sample for training a neural network to approximate the pricing function behind VIX derivatives. Nonetheless, the reader should be aware that we utilized a high-performance computing (HPC) infrastructure to ensure the timely completion of the training data generation phase. \\

We provide a detailed pseudo-code for the Least Squares Monte Carlo (LSMC) method in Algorithm \ref{algo:LSMC}. \\

\begin{algorithm}
\caption{Least Squares Monte Carlo (LSMC)}
\begin{algorithmic}
    \State \% \ $T$: VIX maturity
    \State \% \ $K$: strike price
    \State \% \ $N^\text{sim}_\text{out}$: number of outer paths
    \State \% \ $N^\text{sim}_\text{inn}$: number of inner paths
    \State \% \ $\dt$: discretization step
    \State \% \ $N^\text{steps}_\text{out}$ = $\frac{T}{\dt}$
    \State \% \ $N^\text{steps}_\text{inn}$ = $\frac{\Delta}{\dt}$
    \State \% \ $M$: state variables
    \State \% \ $N^\text{sub}$: number of outer paths used for regression
    \State \% \ $c$: ridge regularizer
    \State \% \ $d$: max degree of monomials in regression
    \State
    \State \% \ simulate the state variables until maturity $T$ in $N^\text{sim}_\text{out}$ independent repetitions  ($M^{(n)}$)
    \For{$k=1:N^\text{steps}_\text{out}$}
        \State $M_{k \times \dt} = \mathrm{simulte\_onestep}(M_{(k-1) \times \dt}, N=N^\text{sim}_\text{out})$ 
    \EndFor
    \State
    \State \% \ apply nested simulation on a subsample of the outer paths
    \State $\widetilde{VIX}^2$ = zeros($N^\text{sub}$,1)
    \For{$n=1:N^\text{sub}$}
        \State $\Sigma$ = [$\sigma_T^{2(n)}$ zeros($N^\text{sim}_\text{inn}$,$N^\text{steps}_\text{inn}$)]
        \For{$k=1:N^\text{steps}_\text{inn}$}
            \State $M_{T+k \times \dt}$ = simulate\_onestep($M_{T+(k-1) \times \dt}$, $N=$ $N^\text{sim}_\text{inn}$)
            \State $\Sigma(:,k+1) = \sigma^{2(n)}_{T+k \times \dt}$
        \EndFor
        \State $I$ = sum($\Sigma$,2) / ($N^\text{steps}_\text{inn}$ + 1)
        \State $\widetilde{VIX}^2(n)$ = mean($I$)
    \EndFor
    \State $\widetilde{VIX}$ = sqrt($\widetilde{VIX}^2$)
    \State
    \State \% \ choose a basis, value it on $M_T^\text{sub}$ and compute ridge regression coefficients
    \State $X$ = monomials($M_T^\text{sub}$,$d$)
    \State $b$ = ridge($X$,$\widetilde{VIX}$,$c$)
    \State
    \State \% \ extend VIX over the full sample $M_T^\text{full}$
    \State $VIX$ = poly\_fun(monomials($M_T^\text{full}$,$d$),$b$)
    \State
    \State \% \ compute derivative prices
    \State $F$ = mean$(VIX)$
    \State $C$ = mean$($max$(VIX-K,0))$
    \State 
    \State \% \ where 
    \State \% \ monomials($M_T$,$d$) produces all monomials of degree up to $d$ given $M_T$
    \State \% \ ridge($X$,$\widetilde{VIX}$,$c$) regresses 
    $\widetilde{VIX}$ against $X$ with an $L_2$ penalization with constant $c$ 
    \State \% \ poly\_fun(monomials($M_T$,$d$),$b$) organizes monomials($M_T$,$d$) into a polynomial function of coefficients $b$ approximating VIX 
\end{algorithmic}\label{algo:LSMC}
\end{algorithm}

Ridge regularization plays a crucial role in ensuring the stability of the regression coefficients. It comes at no additional cost compared to the classical OLS problem, and a closed-form expression for $b^\star$ is readily available. \cite{GG:21} experiment with richer, more complicated models, including neural networks among the regressors and random forests, in an attempt to improve the reliability of the VIX estimator. Given the numerical evidence documented above, there is no need to resort to such complicated and slower approaches.

\section{Neural network approximation} \label{sec:nn}

While LSMC significantly reduces the computational burden of pricing VIX derivatives, the availability of an efficient numerical scheme is just the first brick toward fast neural calibration to real market data. \\

\begin{figure}[ht!]
    \centering
    \includegraphics[width=0.5\linewidth]{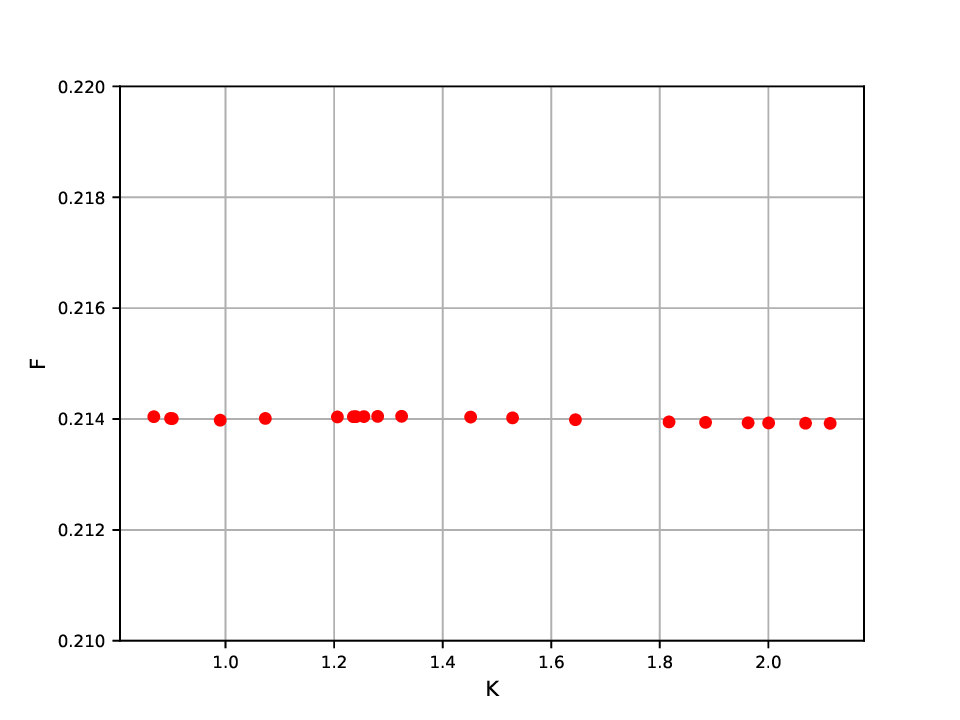}
    \caption[Neural network candidate for the futures price as a function of the strike]{Neural network candidate for the futures price as a function of the strike.}
    \label{fig:F=f(K)}
\end{figure}

As we are examining two markets (SPX and VIX), we require two separate networks. The first one is conceptually very similar to what we have devised in \cite{BBR:24}, mapping model parameters ($\bm{\theta}$) and contract specifications ($T$, $K$) to single SPX implied volatilities (iv for short). We have already introduced this map in the introduction.
The second network handles claims written on the VIX. It outputs the futures and option prices jointly: 
\begin{equation*}
\Phi^{\M,VIX}_p(\bm{\theta},T,K): (\bm{\theta}, T, K) \mapsto 
\left( F^{\M,VIX}(\cdot), C^{\M,VIX}(\cdot) \right).  
\end{equation*}
One key point about the map above is that we leave it to the network to learn that VIX futures prices are independent of the strike input. Figure \ref{fig:F=f(K)} demonstrates that we properly trained the neural network, the neural approximation of the VIX future ($F$) being, within three decimal digits, essentially independent of the strike.

An alternative approach is to learn VIX derivatives through two sub-networks, treating the futures price as a function of model parameters and time to maturity only. However, we claim that jointly learning futures and option prices is beneficial to the training process, as this procedure forces the networks to learn the VIX marginal distribution -- implicit in the option prices -- consistently with its first moment, i.e., the futures price. \\

Learning the pricing maps replaces cumbersome nested Monte Carlo simulations with simple matrix-vector products, reducing calibration time to a few seconds. \\

We now provide all relevant details concerning the network structure and the training process and assess the quality of the neural approximations. 

\subsection{Parameter bounds selection}

The first step towards learning a model in terms of its pricing functions is to identify the relevant region in the parameter space where the price of financial derivatives takes meaningful values from the market perspective. In other terms, we need to restrict the vector below
\begin{equation*}
    \boldsymbol{\theta} = 
    (\underbrace{\beta_0, \beta_1, \beta_2, \beta_{1,2}, 
        \lambda_{1,0}, \lambda_{1,1}, \theta_1, \lambda_{2,0}, \lambda_{2,1}, 
         \theta_2}_{\boldsymbol{\theta}^{\mathcal{M}}},
    \underbrace{R_{1,0,0}, R_{1,1,0}, R_{2,0,0}, R_{2,1,0}}_{\boldsymbol{\theta}^R})
\end{equation*}
to some realistic hypercube in dimension $D=14$. We achieve this via the following multi-step procedure.

\begin{enumerate}
    \item The mixing constants $\theta_1$ and $\theta_2$ have very natural bounds:
    \begin{align*}
        & \theta_1 \in [0,1]\,, \\
        & \theta_2 \in [0,1]\,.
    \end{align*}
    \item The choice of the $\lambda$'s is subtle. We are not considering very short-term options in our exercise, and consequently, we do not need extreme mean-reversion speeds for any of the factors. We therefore assume: 
    \begin{align*}
        & \lambda_{1,0} \in [10,65]\,, \quad \lambda_{1,1} \in [0,35]\,, \\
        & \lambda_{2,0} \in [\hspace{1.9mm} 0,50]\,, \quad \lambda_{2,1} \in [0,15]\,.
    \end{align*}
    \item We start with large conservative bounds for the $\beta$'s.
    \item We uniformly draw about one million ($2^{20}$ to be precise) instances of $\boldsymbol{\theta}^{\mathcal{M}}$ according to the constraints defined above. For each realization, we need to assign initial values to the $R_{n,j}$ Markovian factors. We sample dates uniformly with repetition from the last 15 years of trading, conditionally on the date we collect the historical path of SPX returns, and compute $\boldsymbol{\theta}^R = (R_{1,0,0}, R_{1,1,0}, R_{2,0,0}, R_{2,1,0})$ according to the factor definition.\\
    Once the parameter sampling procedure is complete, we price random surfaces whose maturities extend up to 13 months and automatically inspect the generated smiles. We require that each surface satisfies mild level and shape constraints. If not, we discard it. \\
    The empirical distribution of the surviving surfaces determines the bounds for the initial values of the Markovian factors. Specifically, the choice
    \begin{align*}
        & R_{1,0,0} \in [-1.62,0.88] \quad R_{1,1,0} \in [-1.05,0.71] \\
        & R_{2,0,0} \in [\hspace{7.3mm}0,0.11] \quad R_{2,1,0} \in [\hspace{7.3mm}0,0.11] 
    \end{align*}
    ensures that we exclude no more than 2-5\% of the mass from the tails of the distribution of each factor.
    \item We calibrate a small sample of SPX surfaces (50) and conclude that we can restrict the values of the $\beta$'s to
    \begin{equation*}
        \beta_0 \in [0,0.85]\,, \quad \beta_1 \in [-0.30,-0.10]\,, \quad \beta_2 \in [0.35,0.95]\,, \quad \beta_{1,2} = [0.05,0.40].
    \end{equation*} 
\end{enumerate}

Importantly, while generating training samples, we make sure that 
\begin{align*}
    & \lambda_{1,0} > \lambda_{1,1}\,, \\
    & \lambda_{2,0} > \lambda_{2,1}\,,
\end{align*}
so that the first factor is always the one that mean-reverts faster. This trick prevents the neural network from learning the obvious symmetry that results from exchanging $(\lambda_{\cdot,0},1-\theta_\cdot)$ with $(\lambda_{\cdot,1},\theta_\cdot)$. We impose the same hierarchy on the $\lambda$'s throughout calibration via linear inequality constraints.

\subsection{Learning SPX implied volatilities} 

Learning about SPX implied volatilities proceeds much as we have described in \cite{BBR:24}, with a few minor adjustments that are entirely problem-specific. \\

Firstly -- and most importantly -- model parameters are not sampled uniformly from the parameter space, at least not all of them. We just stressed that ensuring a certain hierarchy $\lambda_{\cdot,0} > \lambda_{\cdot,1}$ is essential for reasons of symmetry. The key point is that the factors' initial values are inherently linked to the $\lambda$'s via the history of the index. More specifically, we have 
\begin{equation*}
    R_{n,j,0} = \lambda_{n,j} \sum_{i=0}^{T-2} \e^{-\lambda_{n,j} \frac{i}{252}} \left(\frac{S_{-i}}{S_{-i-1}}-1
    \right)^n \qquad n = \{ 1,2 \}, \quad j = \{ 0, 1\}
\end{equation*}
where $T = 1008$ selects a 4-year price path starting at a random date between January 1, 2009, and December 31, 2023 (with value $S_0$). \\
The result is an empirical distribution of the $R_{n,j,0}$'s that is far from uniform. \\

Moreover, only a small part of the parameter combinations we have drawn enter the training sample. A lot of them correspond to completely unrealistic surfaces that we discard based on two simple conditions:
\begin{itemize}
    \item the left-most point in the left wing of the smile (name it $\sigma^{w_l}$) is such that $\sigma^{w_l} < 0.60$, for each smile in the surface; 
    \item the ratio of the left-most point in the left wing to the right-most point in the right wing satisfies $\frac{\sigma^{w_l}}{\sigma^{w_r}}<1.50$, for each smile in the surface.
\end{itemize}

The purpose of the shape constraints above is to ensure that the network only learns smiles similar to those we encounter in the real market. However, such a seemingly reasonable practice has a significant drawback from a calibration perspective. During the calibration process, the optimizer may explore regions corresponding to parameter values not sampled during the training phase. The networks will output inaccurate implied volatilities, thus misleading the gradient for the next step. To remedy this, we add a uniform buffer of parameters to the training sample, not subject to any validation. If the buffer is sufficiently large, it should facilitate learning and enhance robustness. At the same time, the buffer cannot be too large if we want the neural approximation to work appropriately on the typical smiles. Therefore, we suggest that it accounts for about 15-20\% of the whole training sample. \\

\begin{figure}[!ht]
    \centering
    \includegraphics[trim={0cm 4cm 0cm 8cm},clip,width=\linewidth]{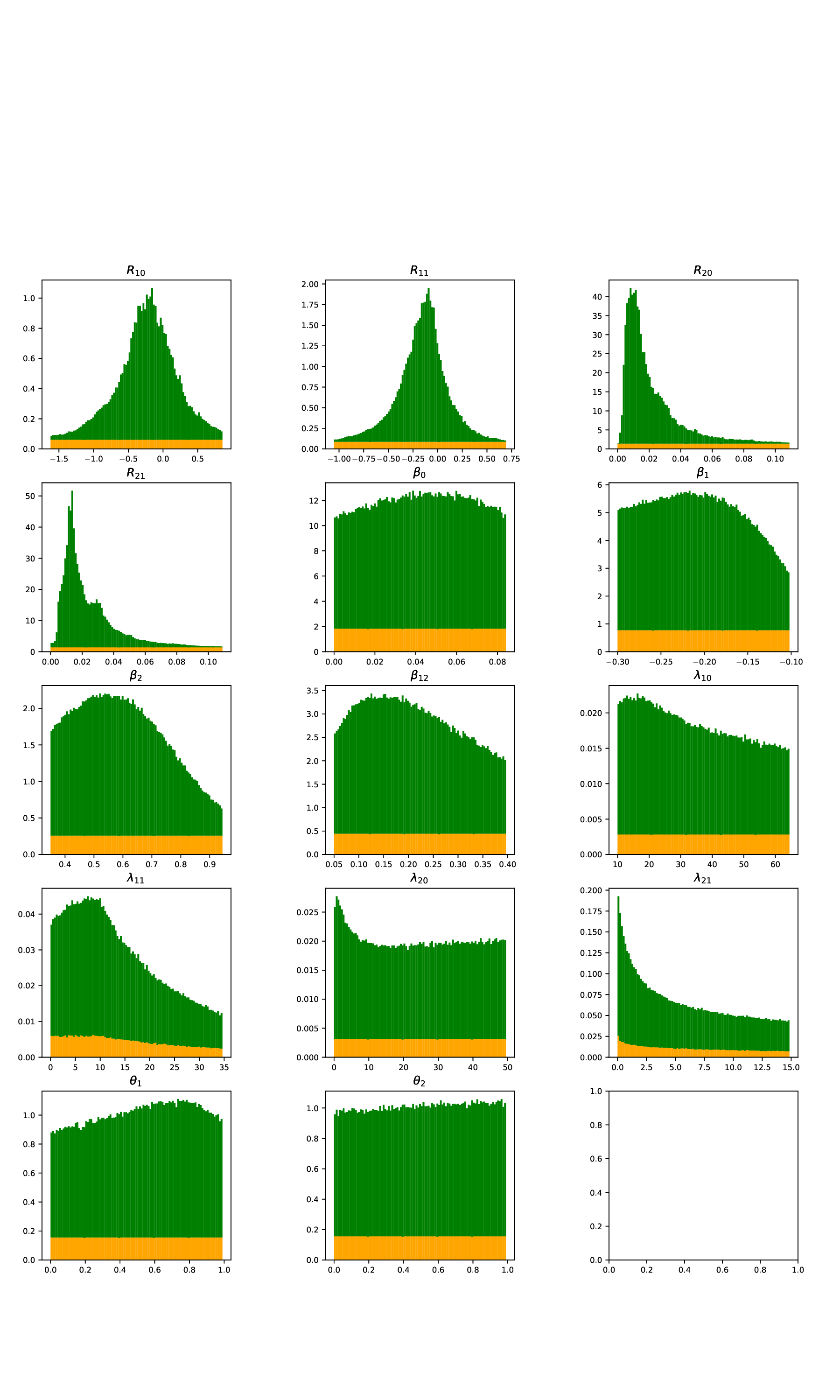}
    \caption[Histograms of the model parameters as sampled for the training of a neural network for SPX implied volatilities.]{Histograms of the model parameters as sampled for the training of a neural network for SPX implied volatilities. Uniform buffer in orange. Shape-constrained sampling in green.}
    \label{fig:histo_spx_train}
\end{figure}

Finally, except for the inequality constraints on the $\lambda$'s, the buffer is purely uniform, ensuring that each region in the hypercube defining the parameter space is adequately populated. Histograms for all the parameters are shown in Figure \ref{fig:histo_spx_train}. The uniform buffer in orange represents approximately 15\% of 400,000 parameter sets. \\

Each $\bm{\theta} = (\bm{\theta}^{\mathcal{M}}, \bm{\theta}^R)$ in the training sample is associated with a random grid of times to maturity and strikes. We sample 11 expirations uniformly at random from inside the time sub-intervals in $[T_{\min}, T_{\max}]$ below:

\begin{align*}
[T_{\min},T_{\max}] = & [6/365,1/12) \cup [1/12,2/12) \cup [2/12,3/12) \cup [3/12,4/12) \cup [4/12,5/12) \cup \nonumber \\ & [5/12,6/12) \cup [6/12,8/12) \cup [8/12,10/12) \cup [10/12,11/12) \cup [11/12,1) \cup \nonumber \\ & [1,13/12].
\end{align*}

Strikes are also random, but they extend around the ATM according to a simple square-root rule 
\begin{align}\label{eqn:K_sqrt_bounds}
        [K_{\min}(T),K_{\max}(T)] = [S_0(1-l\sqrt{T}),S_0(1+u\sqrt{T})], \qquad l=0.55 \quad u=0.30.
\end{align}
Additionally, we distinguish three levels of moneyness $\left(\frac{K}{S_0}\right)$, where we sample strikes with varying granularities. We refer to \cite{BBR:24} for all the details. \\

We fill the grid buckets using Monte Carlo simulation with $N_{\mathrm{sim}}=2^{18}$ trajectories. This last step concludes the generation of the training set. \\

A simple feed-forward neural network learns Black-Scholes implied volatilities as a function of model parameters $\bm{\theta}$ and contract specification $(T, K)$. \\
We use five hidden layers, each with 128 nodes, and employ the swish activation function. The output layer is linear. The RMSE measures the loss, and the Adam optimizer runs for at most 500 epochs (unless early stopping occurs). We teach the network to restore the best set of weights it met during training and add no other condition. The MinMaxScaler scales the input; we do not scale the output. Keras is our preferred engine. \\

\begin{figure}[!ht]
    \centering
    \includegraphics[width=0.5\linewidth]{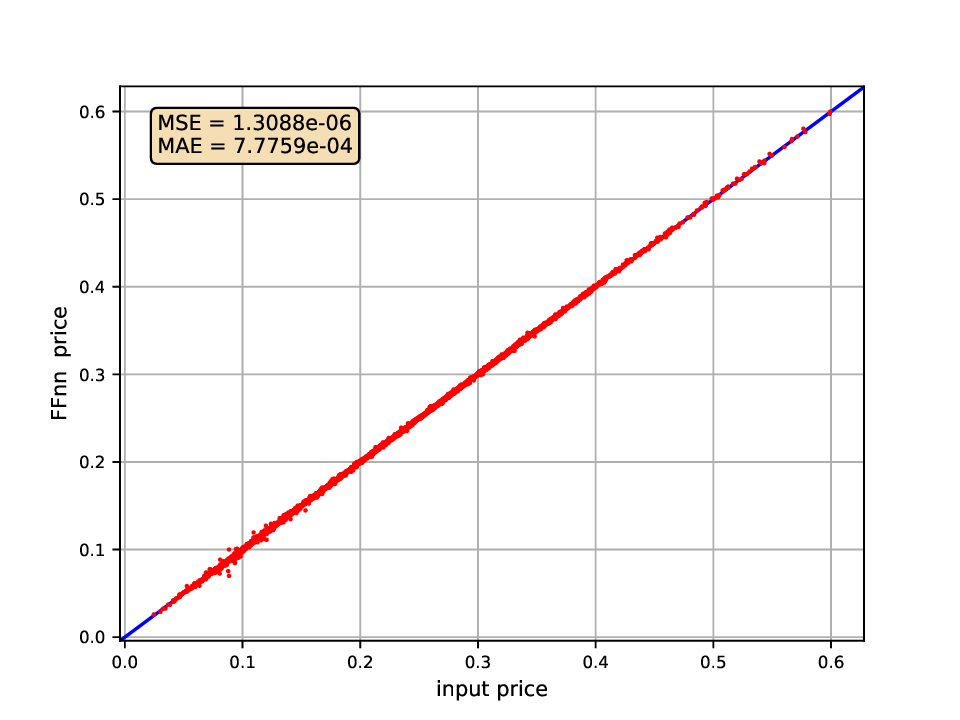}
    \caption[Out-of-sample performance of the neural approximation for SPX implied volatilities.]{Out-of-sample performance of the neural approximation for SPX implied volatilities. Mean squared error (MSE) and mean absolute error (MAE) in the box.}
    \label{fig:v_spx_deg45_oos}
\end{figure}

Figure \ref{fig:v_spx_deg45_oos} summarizes the out-of-sample performance of the network. We do not include any surface failing to meet the shape conditions in the test set. That is because our primary concern is that the neural approximation works fine on the typical smiles. 

The 45-degree line leaves little space for interpretation. The agreement between the neural approximation and the target is excellent, and the mean absolute error (MAE) indicates that we reproduce the correct volatility to an average of 3 decimal digits. \\

We store the optimal neural network weights on disk and transfer everything to MATLAB for pricing and calibration. We will see the network in action in the numerical section. 

\subsection{Learning VIX derivative pricing}

We use a blended approach for learning the price of VIX derivatives (options and futures) as well. The relative proportion of the uniform buffer is the same as before, but we do not enforce any shape condition on the smiles. \\

The fact that the target is now a price is extremely natural for at least two reasons: 1) there is no such thing on the market as an implied volatility for futures, and 2) the implied volatility of a VIX option depends on the VIX futures, while the price of a VIX option does not. Then, learning implied volatilities directly would be affected by two sources of statistical uncertainty: the Monte Carlo computation of the option price and the computation of the futures price. \\

As already mentioned, we learn VIX options and futures together (i.e., with one unique network). We consider about a million random grids in the training set, each of which consists of two maturities in
\begin{align*}
[T_{\min},T_{\max}] = & [6/365,18/365) \cup [18/365,30/365) 
\end{align*}
and 40 strikes in total. Strikes are quoted in terms of the futures and belong to
\begin{align*}
    m := \frac{K}{F} = [0.82,2.36]
\end{align*}
irrespectively of the expiration. As a consolidated practice, we distinguish three levels of moneyness and take
\begin{itemize}
    \item 4 strikes in $[0.82,1.00]$ 
    \item 7 strikes in $[1.00,1.40]$
    \item 9 strikes in $[1.40,2.36]$.
\end{itemize}

Rows in the training sample are therefore given as
\begin{equation*}(\bm{\theta},T_1,T_2,m_1^1,\dots,m_1^{20},m_2^1,\dots,m_2^{20},F_1,F_2,C_1^1,\dots,C_1^{20},C_2^1,\dots,C_2^{20}).
\end{equation*}

These are the result of a time-consuming generation phase. Each surface takes approximately 7 seconds to process on a single node with 48 CPU cores. We use up to 100 nodes in parallel. The times we quote refer to code written in C language leveraging the LSMC with $N_{\mathrm{sim\_out}}=2^{18}$, $N_{\mathrm{sample}} = 2^{13}$, $N_{\mathrm{sim\_inn}} = 2^{10}$. \\

\begin{figure}[!ht]
\centering
\begin{tikzpicture}[x=2.3cm,y=1cm]
  \readlist\Nnod{3,5,5,5,5,5,2} 
  \readlist\Nstr{3,256,256,256,256,256,2}
  \readlist\Cstr{\strut x,h^{(\prev)},h^{(\prev)},h^{(\prev)},h^{(\prev)},h^{(\prev)},y}
  \message{^^J  Layer}
  \foreachitem \N \in \Nnod{ 
    \edef\lay{\Ncnt} 
    \message{\lay,}
    \pgfmathsetmacro\prev{int(\Ncnt-1)} 
    \foreach \i [evaluate={\y=\N/2-\i; \index=(\i<\N?int(\i):"\Nstr[\lay]"); \x=\lay; \n=\nstyle;}] in {1,...,\N}{ 
      
      \ifnum\lay=1 
        \ifnum\i=1
          \node[node \n] (N\lay-\i) at (\x,\y) {$\bm{\theta}$}; 
        \fi
        \ifnum \i=2
          \node[node \n] (N\lay-\i) at (\x,\y) {$T$}; 
        \fi
        \ifnum \i=3
          \node[node \n] (N\lay-\i) at (\x,\y) {$m$}; 
        \fi
      \else
        \ifnum\lay=7
          \ifnum\i=1
            \node[node \n] (N\lay-\i) at (\x,\y) {$F$};
          \fi
          \ifnum\i=2
            \node[node \n] (N\lay-\i) at (\x,\y) {$C$};
          \fi
        \else
          \ifnum \i=4
            \node[node \n] (N\lay-\i) at (\x,\y) {$\vdots$}; 
          \else 
            \node[node \n] (N\lay-\i) at (\x,\y) {$\Cstr[\lay]_{\index}$};
          \fi
        \fi
     \fi
      
      \ifnum\lay>1 
        \foreach \j in {1,...,\Nnod[\prev]}{ 
          \draw[connect arrow] (N\prev-\j) -- (N\lay-\i); 
        }
      \fi 
      
    }
    
  }

  \node[above=35.5,align=center] at (N1-1.90) {$\mathrm{input}$};
  \node[above=8,align=center] at (N2-1.90) {$\mathrm{hidden 1}$};
  \node[above=8,align=center] at (N3-1.90) {$\mathrm{hidden 2}$};
  \node[above=8,align=center] at (N4-1.90) {$\mathrm{hidden 3}$};
  \node[above=8,align=center] at (N5-1.90) {$\mathrm{hidden 4}$};
  \node[above=8,align=center] at (N6-1.90) {$\mathrm{hidden 5}$};
  \node[above=50,align=center] at (N7-1.90) {$\mathrm{output}$};
\end{tikzpicture}
\caption[Neural network architecture for VIX derivatives]{Neural network architecture for VIX derivatives. The input layer is $p+2$-dimensional, where $p=14$ is the number of model parameters. The output layer is 2-dimensional. $F$ is the future, $C$ is the call option price.}
\label{fig:net_vix}
\end{figure}
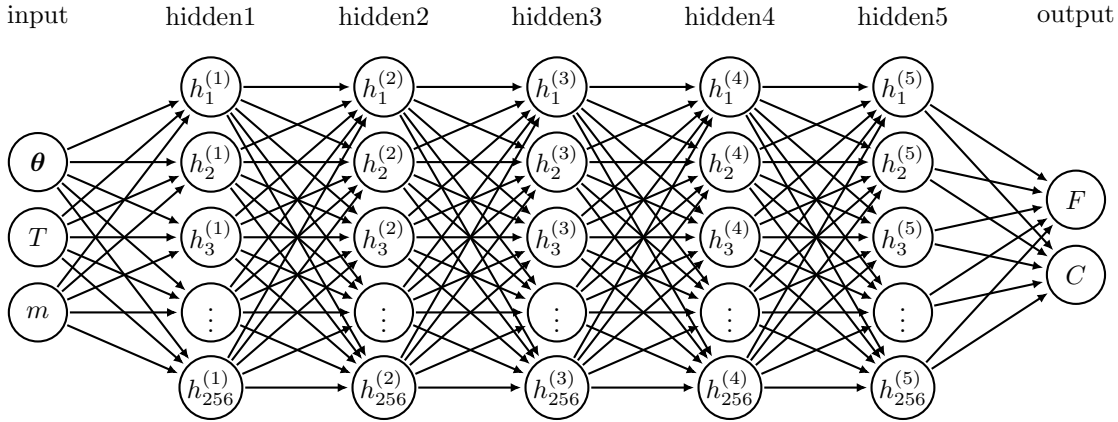

We flatten samples for pointwise training and pass them to a feed-forward neural network whose structure is described in Figure \ref{fig:net_vix}.

We now have five hidden layers, each with 256 nodes, and utilize the swish activation function. The output layer is linear. The MSE measures the loss, and the Adam optimizer runs for at most 1000 epochs (unless early stopping occurs). We teach the network to restore the best set of weights it met during training and add no other condition. The MinMaxScaler scales the input, and the output uses a StandardScaler. \\

\begin{figure}[!ht]
    \centering
    \begin{subfigure}[b]{0.49\textwidth}
        \centering
        \includegraphics[width=\textwidth]{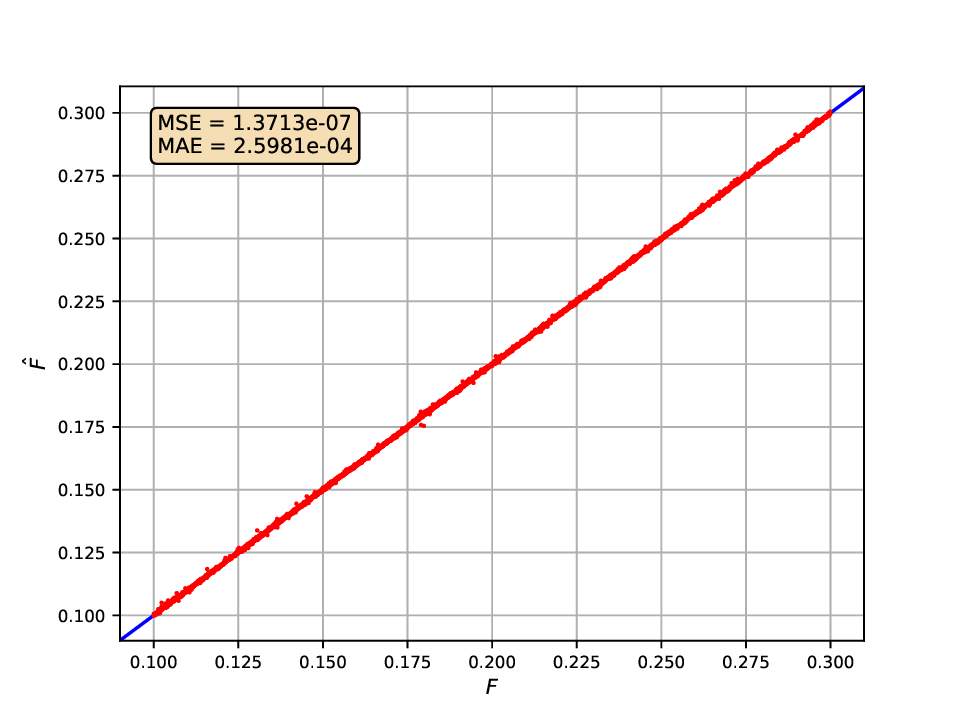}
    \end{subfigure}
    \begin{subfigure}[b]{0.49\textwidth}
        \centering
        \includegraphics[width=\textwidth]{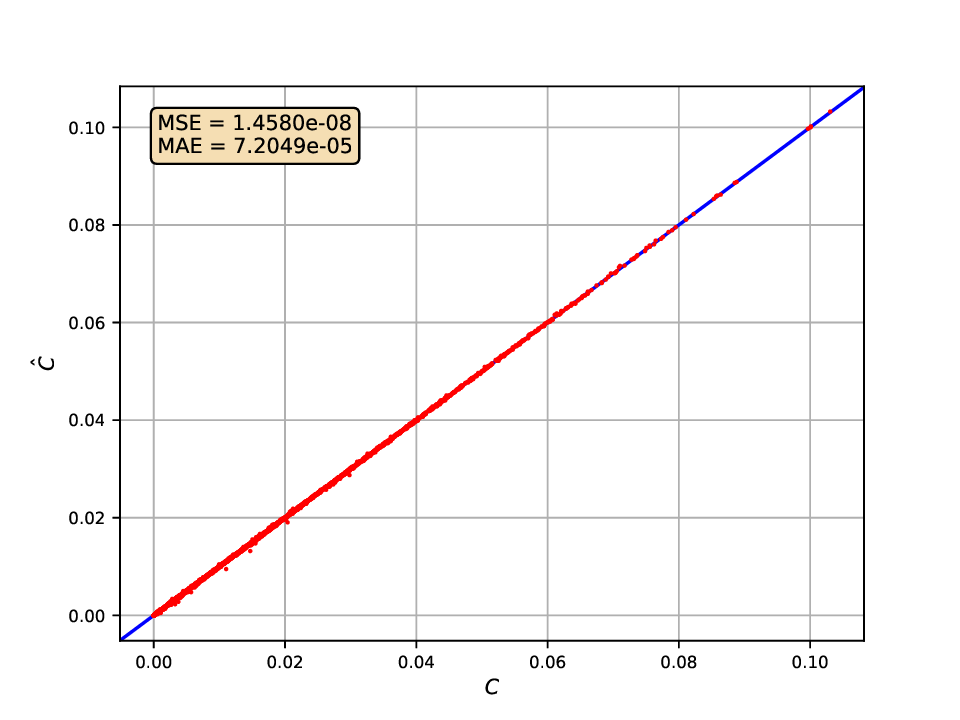}
    \end{subfigure}
    \caption[Out-of-sample performance of the neural approximation for VIX futures (left) and VIX call option prices (right).]{Out-of-sample performance of the neural approximation for VIX futures (left) and VIX call option prices (right). Mean squared error (MSE) and mean absolute error (MAE) in the box.}
    \label{fig:nn_vix_45deg}
\end{figure}

We investigate the quality of the neural approximation in terms of its performance for futures and call option prices separately. The test set only consists of points in the parameter space where the future is such that $F \in [0.10, 0.30]$. Again, the idea is that this is what we typically find in the market, and we want to learn it as accurately as possible. If the condition on the futures holds, then we test the associated calls. \\
Figure \ref{fig:nn_vix_45deg} is the usual 45-degree lines: call option prices on the right, futures on the left. We add nothing more about the calls but are particularly interested in the error associated with the futures. This price is indeed a variable in the Black formula when inverting for VIX implied volatilities. One needs to learn it with extreme precision. We are reassured to see that the MAE is of the same magnitude as the Monte Carlo error in the target.

\begin{figure}[!ht]
    \centering
    \begin{subfigure}[b]{0.49\textwidth}
        \centering
        \includegraphics[width=\textwidth]{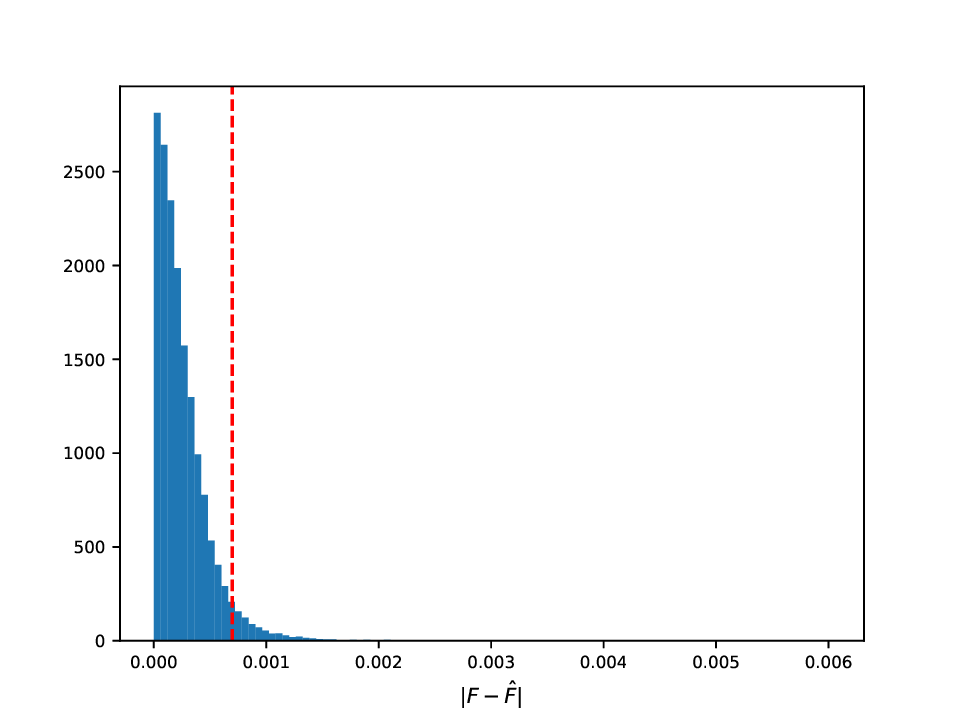}
    \end{subfigure}
    \begin{subfigure}[b]{0.49\textwidth}
        \centering
        \includegraphics[width=\textwidth]{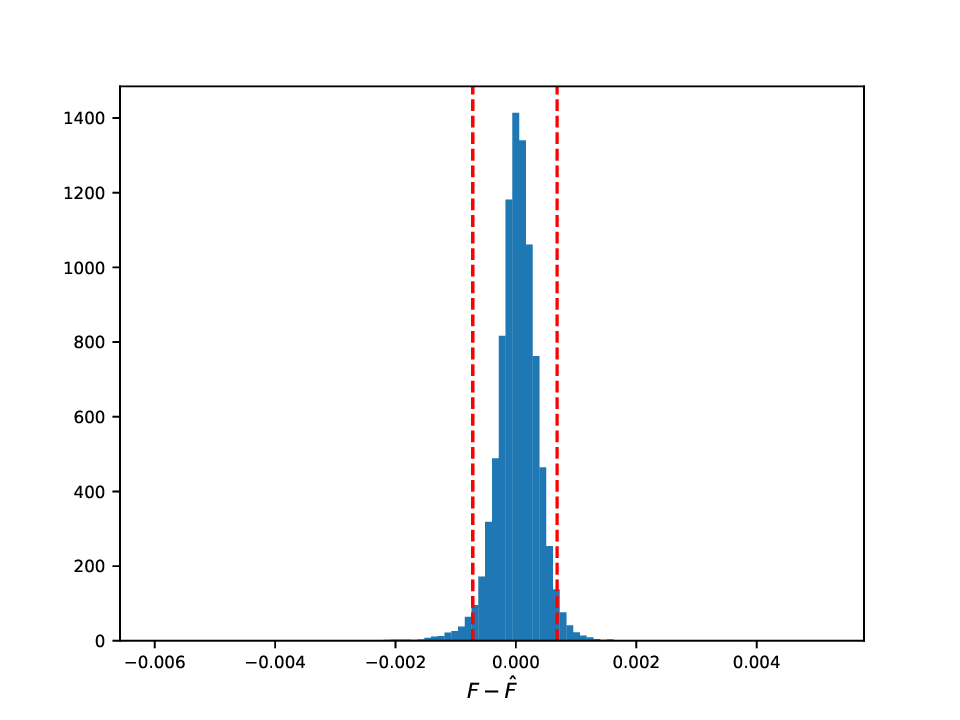}
    \end{subfigure}
    \caption[Out-of-sample error distribution of the VIX futures (1- and 2-tailed).]{Out-of-sample error distribution of the VIX futures (1- and 2-tailed). 95\% quantiles superimposed in dashed red: $q_{0.950}^{|\mathrm{err}|}=6.9774\e-04$, $q_{0.025}^{\mathrm{err}}= -7.1743\e-04$, $q_{0.975}^{\mathrm{err}}= 6.8453\e-04$.} 
    \label{fig:nn_vix_F_histo}
\end{figure}

Figure \ref{fig:nn_vix_F_histo} demonstrates that the absolute error is lower than $7\e-04$ in 95\% of the cases. We conclude that our approximation of the future is excellent.

\section{Calibration} \label{sec:calib}

In this Section, we describe the calibration process, the structure of the loss function, and the empirical results, both for the single SPX problem and the joint one. \\

Following the example by \cite{GG:25}, we optimize only over parameters $\bm{\theta}^{\mathcal{M}} \in \R^{10}$. So, at every iteration in the calibration loop, and for every $n=\{1,2\}$ and $j=\{0,1\}$, we evaluate the $R_{n,j,0}$'s (initial condition on the factors) based on the current value of $\lambda_{n,j}$ and the history of the SPX (4 years from the calibration date) to build $\bm{\theta}^R$. We stack $\bm{\theta} = (\bm{\theta}^{\mathcal{M}},\bm{\theta}^R)$ and pass it to the neural network(s) for evaluation of the loss and its gradient. We take one step and repeat until convergence. \\   

Then, strictly speaking, we do not calibrate $\bm{\theta}^R$, but the optimal $\bm{\theta}^{\mathcal{M}}$ actually implies its value. \\

As for the loss functions, calibration of the SPX surface leverages on the $L_2$ distance between the market volatilities and their model counterpart:
\begin{equation*}
    L^S = \frac{1}{N_T^{SPX}} \sum_{i=1}^{N_T^{SPX}} \left[ \frac{1}{N_{K(T_i^{SPX})}} \sum_{l=1}^{N_{K(T_i^{SPX})}} \left( \sigma_{mkt}^{SPX}(T_i^{SPX},K_l(T_i^{SPX})) - \sigma_{mdl}^{SPX}(T_i^{SPX},K_l(T_i^{SPX})) \right)^2 \right].
\end{equation*}

Things are much more involved for the joint problem as we have to account for
\begin{enumerate}
    \item the contribution from the SPX smiles,
    \item the contribution from VIX futures,
    \item the contribution from the VIX smiles.
\end{enumerate}
We opt for the following
\begin{align}\label{eqn:loss_joint}
    L^J & = w_{vSPX} \frac{1}{N_T^{SPX}} \sum_{i=1}^{N_T^{SPX}} \left[ \frac{1}{N_{K(T_i^{SPX})}} \sum_{l=1}^{N_{K(T_i^{SPX})}} \left( \frac{\sigma_{mdl}^{SPX}(T_i^{SPX},K_l(T_i^{SPX}))}{\sigma_{mkt}^{SPX}(T_i^{SPX},K_l(T_i^{SPX}))} -1 \right)^2 \right] \ + \nonumber \\
      & + w_{fVIX} \frac{1}{N_T^{VIX}} \sum_{i=1}^{N_T^{VIX}} \left( \frac{F_{mdl}^{VIX}(T_i^{VIX})}{F_{mkt}^{VIX}(T_i^{VIX})} - 1\right)^2 \ + \nonumber \\
      & + w_{vVIX} \frac{1}{N_T^{VIX}} \sum_{i=1}^{N_T^{VIX}} \left[ \frac{1}{N_{K(T_i^{VIX})}} \sum_{l=1}^{N_{K(T_i^{VIX})}} \left( \frac{\sigma_{mdl}^{VIX}(T_i^{VIX},F_{mdl}^{VIX}(T_i^{VIX}),K_l(T_i^{VIX}))}{\sigma_{mkt}^{VIX}(T_i^{VIX},F_{mkt}^{VIX}(T_i^{VIX}),K_l(T_i^{VIX}))} -1 \right)^2 \right],
\end{align}
where 
\begin{itemize}
    \item $w_{vSPX}, \ w_{fVIX}, \ w_{vVIX} \in \R$ weight the three contributions above, 
    \item $N_T^{SPX/VIX}$ is the number of SPX/VIX maturities,
    \item $N_{K(T_\cdot^{SPX/VIX})}$ is the number of strikes for a given SPX/VIX maturity.
\end{itemize}

The accuracy we have documented for the VIX futures is crucial for a reliable inversion of the call option price. \\

Equation \eqref{eqn:loss_joint} is very similar to the one proposed by \cite{GM:22}. The only difference is that they calibrate on VIX call option prices instead and consequently avoid inversion from a noisy futures price (which they compute by Monte Carlo integration while simulating neural SDEs). \\

\begin{remark}
    VIX strikes in Equation \eqref{eqn:loss_joint} are normalized at the level of the futures. We should think of them as $K^{VIX} = mF^{VIX}$, where $m$ is the same moneyness as in the inputs of the network. Accommodating for this change of variables is trivial.
\end{remark}

Before we proceed to the result section, it is worth pointing out one last thing. The neural network used for the index in the joint problem differs from the one used for the single problem. There is documented evidence that the 4-factor Markov PDV model is not flexible enough to calibrate SPX maturities longer than a couple of months jointly with VIX smiles. Then, we train a new network to learn short SPX maturities specifically. 

\subsection{The single problem}

\begin{figure}[p]
    \centering
    \begin{subfigure}[b]{0.46\textwidth}
        \centering
        \includegraphics[width=\textwidth]{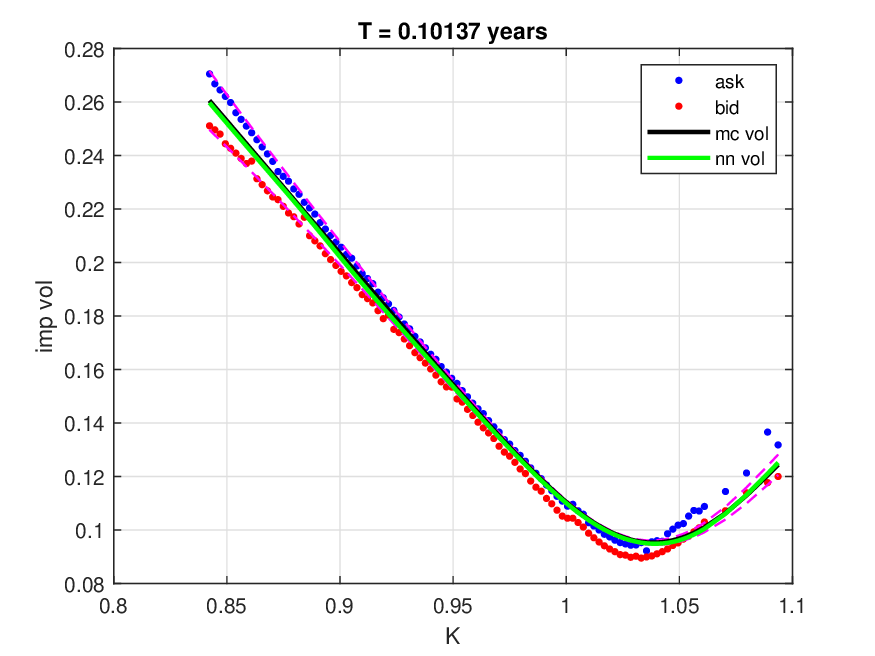}
    \end{subfigure}
    \begin{subfigure}[b]{0.46\textwidth}
        \centering
        \includegraphics[width=\textwidth]{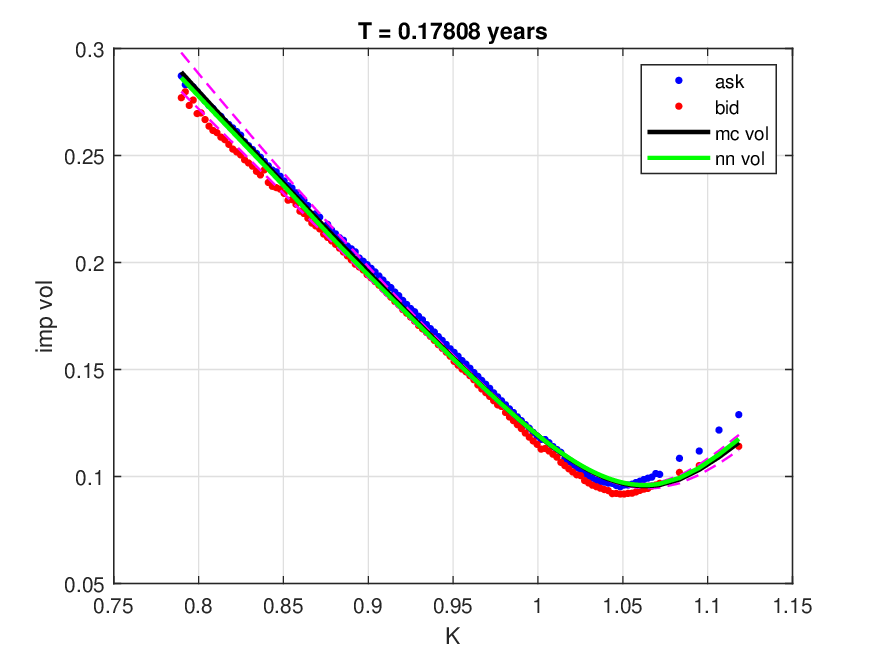}
    \end{subfigure}
    \begin{subfigure}[b]{0.46\textwidth}
        \centering
        \includegraphics[width=\textwidth]{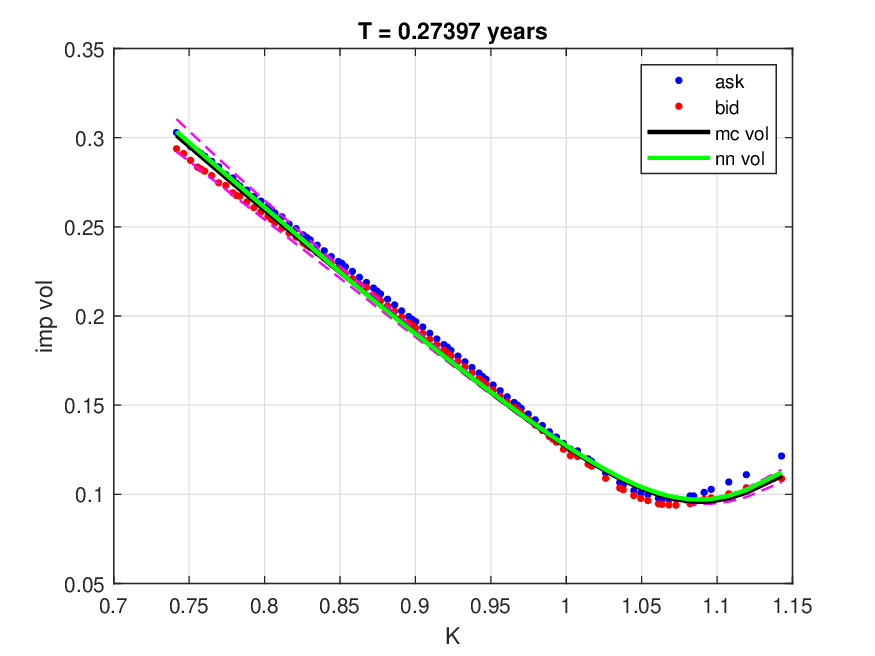}
    \end{subfigure}
    \begin{subfigure}[b]{0.46\textwidth}
        \centering
        \includegraphics[width=\textwidth]{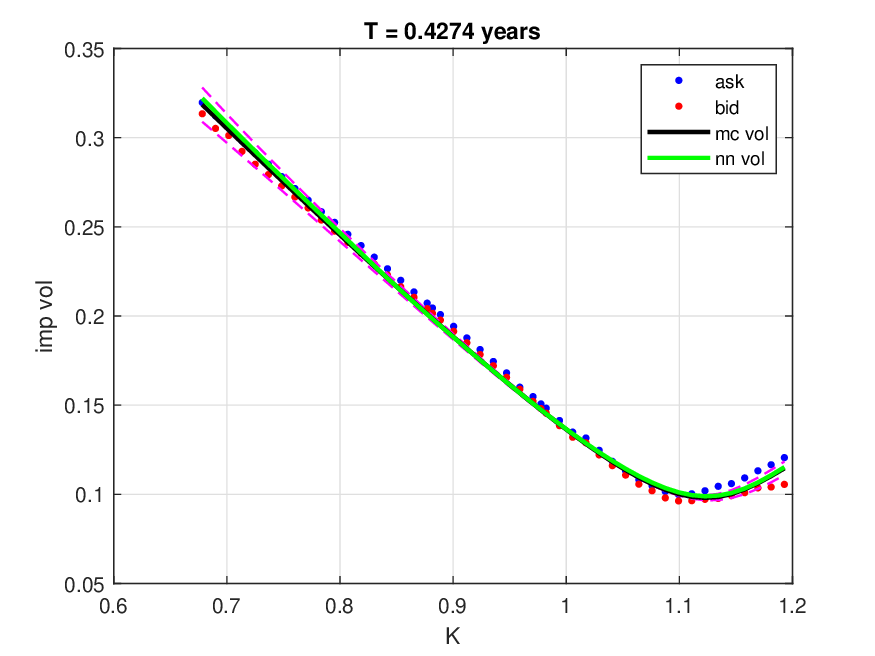}
    \end{subfigure}
    \begin{subfigure}[b]{0.46\textwidth}
        \centering
        \includegraphics[width=\textwidth]{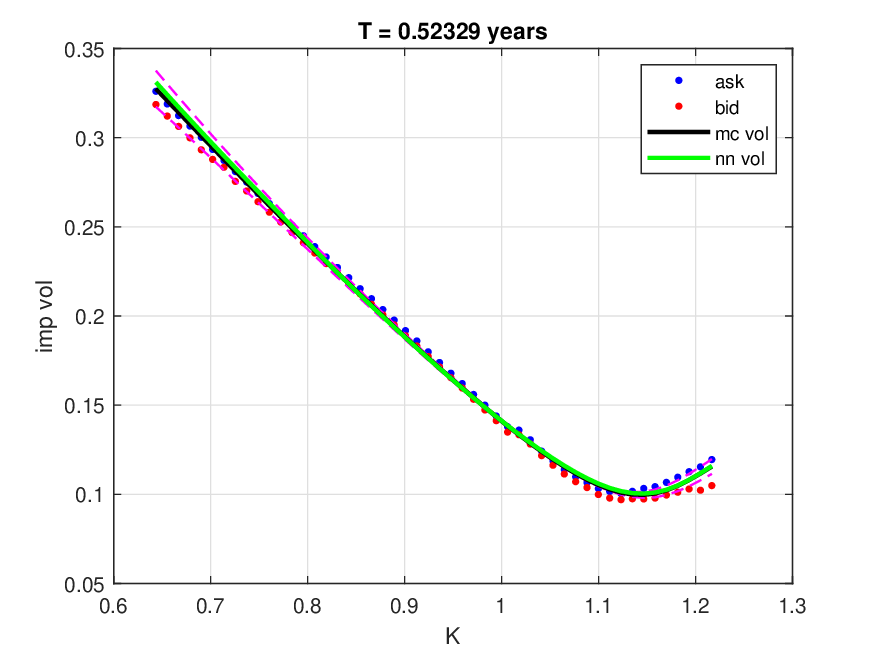}
    \end{subfigure}
    \begin{subfigure}[b]{0.46\textwidth}
        \centering
        \includegraphics[width=\textwidth]{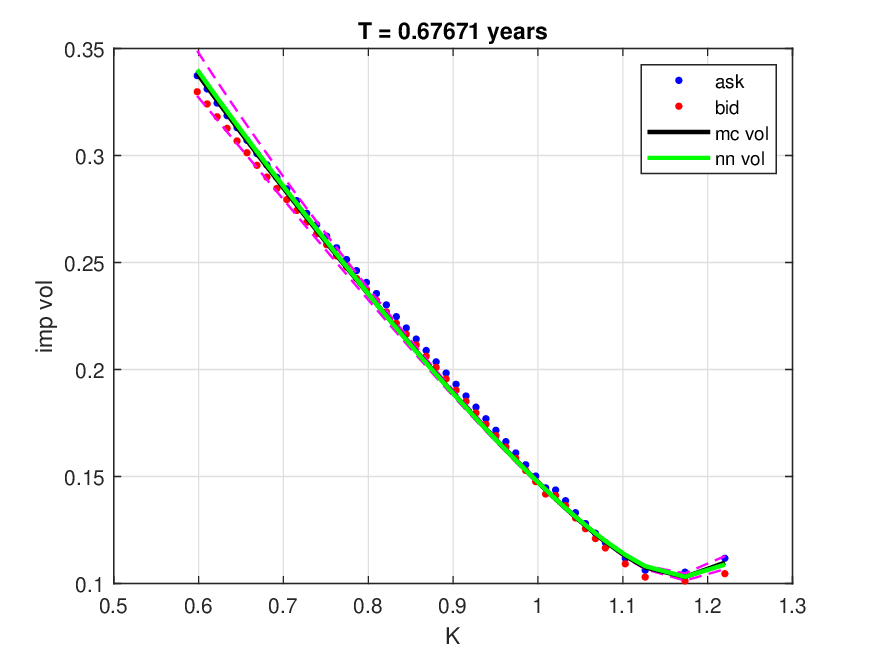}
    \end{subfigure}
    \begin{subfigure}[b]{0.46\textwidth}
        \centering
        \includegraphics[width=\textwidth]{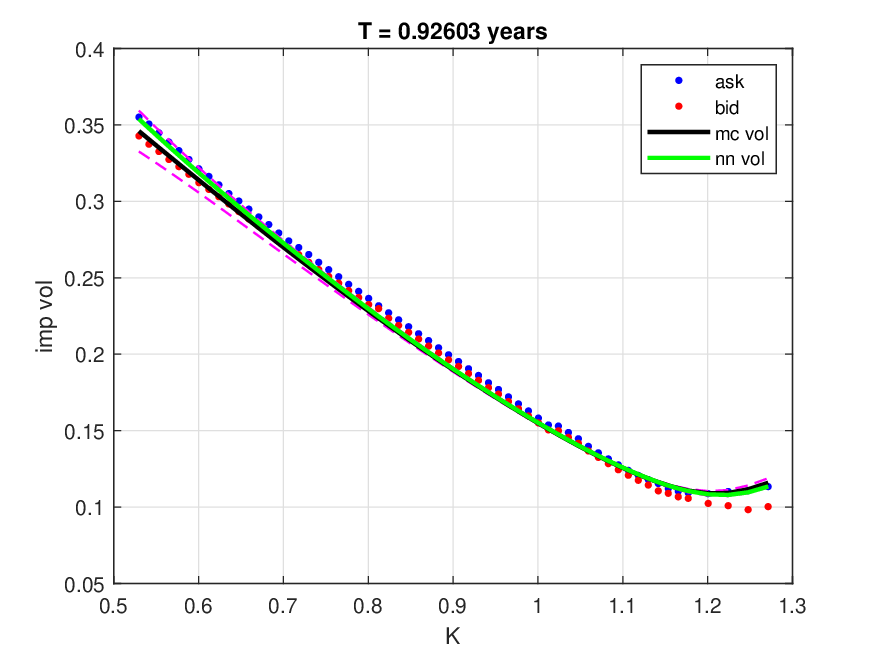}
    \end{subfigure}
    \begin{subfigure}[b]{0.46\textwidth}
        \centering
        \includegraphics[width=\textwidth]{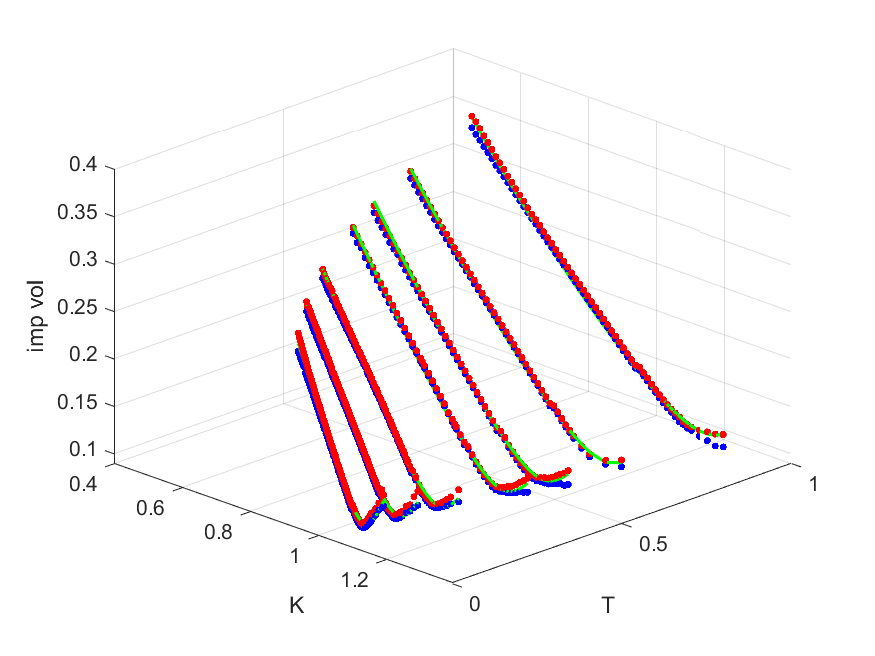}
    \end{subfigure}
    \caption[SPX calibration of the 4-factor Markov PDV model as of July 13, 2016. Fits.]{SPX calibration of the 4-factor Markov PDV model as of July 13, 2016. The neural network fit in green. Monte Carlo benchmarks (together with 95\% confidence bands) in (dashed) black. A comprehensive view of the surface in the bottom right corner.} 
    \label{fig:spx_fits_20160713}
\end{figure}

We only look at one surface in full detail. The quality of the (green) neural network fit in Figure \ref{fig:spx_fits_20160713} is excellent, and all smiles perfectly align with the high-precision Monte Carlo benchmark (within the 95\% confidence bands in dashed black). \\

We report the optimal parameters in Table \ref{tab:spx_params_20160713}, together with the initial values of the factors. These results present an opportunity to emphasize a crucial point. The way we calibrate has no mechanism to prevent the $R_{n,j,0}$'s from escaping the training domain. This particular event occurred on July 13, 2016 (with the first component of $R_1$ exceeding its upper bound). Till, the network extrapolates very well, as demonstrated by the excellent correspondence with the benchmark. \\

Because we have no theoretical guarantee that extrapolation always works fine, it is still important to limit similar occurrences. Our choice to bound the initial values of the factors based on the quantiles of their distributions effectively aligns with this direction. \\

However, there is at least an alternative. The presence of a uniform buffer during generation and training also serves the purpose of disentangling the initial values of the factors from their mean-reversion speed $\lambda_{n,j}$. Then, we could break this relation during calibration as well. In other words, we could optimize in 14-dimensional space and calibrate the boundary conditions $\bm{\theta^R}$ together with $\bm{\theta}^{\mathcal{M}}$, subject to the same box constraints used during generation. This approach should provide additional flexibility for calibration and yield better fits. Nonetheless, such a convenient practice is not in line with the spirit of the model, as it ignores the history of the SPX rather than utilizing the true market path. 

\subsubsection{July 13, 2016}

\begin{table}[!ht]
    \centering
    \begin{tabular}{|c|c|c|c|c|c|c|c|c|c|}
        \hline
        $\beta_0$ & $\beta_1$ & $\beta_2$ & $\beta_{1,2}$ & $\lambda_{1,0}$ & $\lambda_{1,1}$ & $\theta_1$ & $\lambda_{2,0}$ & $\lambda_{2,1}$ & $\theta_2$ \\ \hline
        0.0834 & -0.2427 & 0.3500 & 0.3047 & 59.31 & 7.50 & 0.6692 & 30.13 & 6.55 & 1.0000 \\ \hline
    \end{tabular}
    \caption[SPX calibration of the 4-factor Markov PDV model as of July 13, 2016. Parameters.]{SPX calibration of the 4-factor Markov PDV model as of July 13, 2016. Model parameters as above. Initial values of the factors: $R_{1,0,0}=1.0856$, $R_{1,1,0}=0.2947$, $R_{2,0,0}=0.0298$, $R_{2,1,0}=0.0234$.}
    \label{tab:spx_params_20160713}
\end{table}

\subsubsection{A time-series of MSEs and MAEs}

We now aim to conduct a comprehensive empirical assessment of the 4-factor Markov PDV model, calibrated to SPX market data from 2010 to 2023. For this, we compare it with rough volatility models, namely the rHeston model and the rBergomi model. \\

\begin{figure}[!ht]
    \centering
    \begin{subfigure}[b]{0.49\textwidth}
        \centering
        \includegraphics[width=\textwidth]{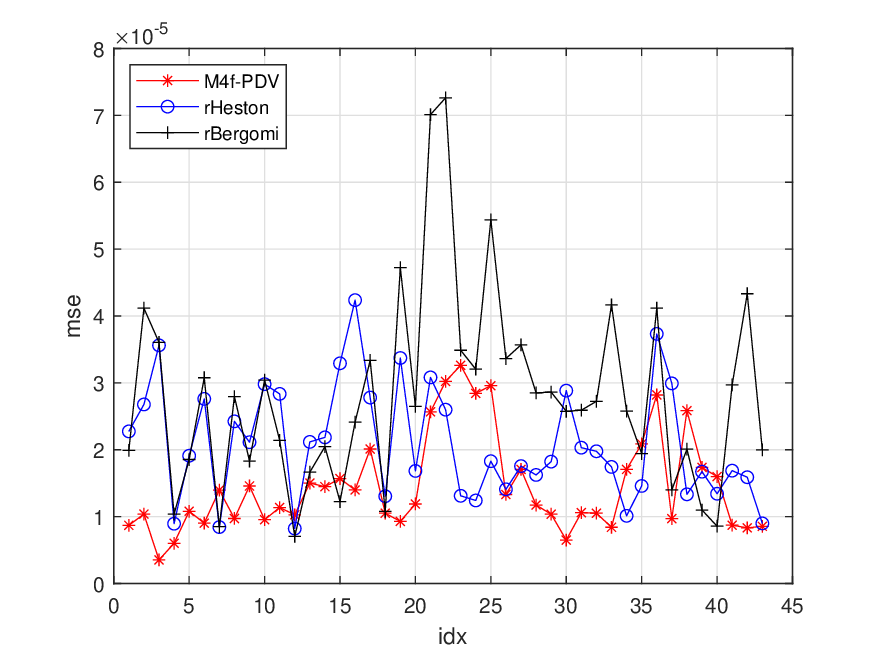}
    \end{subfigure}
    \begin{subfigure}[b]{0.49\textwidth}
        \centering
        \includegraphics[width=\textwidth]{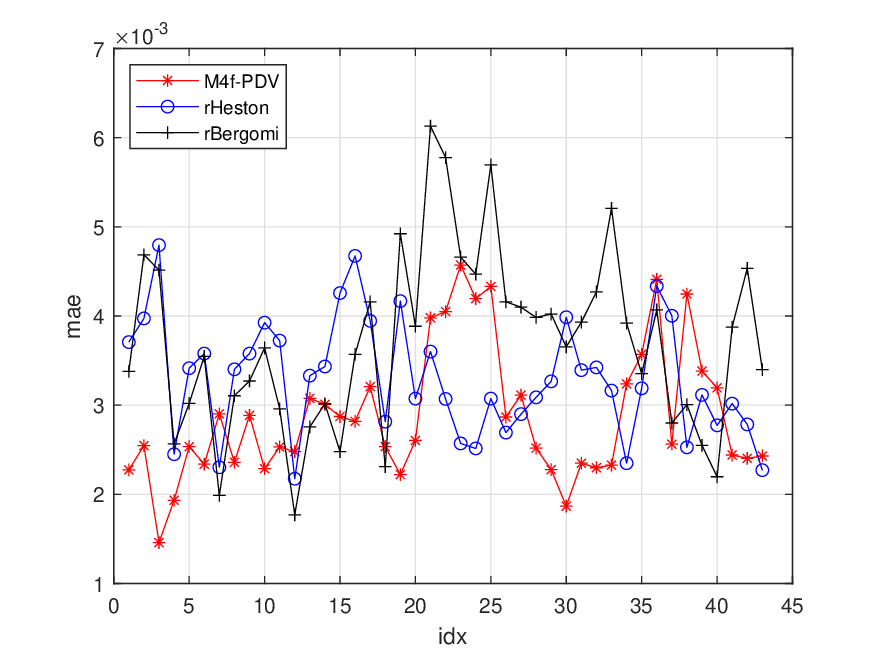}
    \end{subfigure}
    \caption{MSE (left) and MAE (right) over 43 SPX volatility surfaces between 2010 and 2023. Comparing the 4-factor Markov PDV model with rough instances:  the rHeston and the rBergomi models.}
    \label{fig:spx_calib_err_models_ts}
\end{figure}

Figure \ref{fig:spx_calib_err_models_ts} demonstrates that the 4-factor Markov PDV model outperforms its rough competitors (both in terms of MSE and MAE) in almost all the market dates we consider. This fact is not surprising as we now have 10 parameters addressing the static properties of the underlying (i.e., the shape of the smile) as opposed to 3 (Hurst exponent, volatility of volatility, and spot-volatility correlation). Still, rHeston and rBergomi take advantage of an entire curve (either parametric or not) to correct the level of the smiles over time. A flexibility that the 4-factor Markov PDV model is missing. Such a lack of flexibility is typically not an issue when examining monthly maturities -- as we are doing here. It becomes a limitation as soon as one focuses on short-maturity options, where the term structure of the ATM volatility may be very irregular. The absence of a shift is also detrimental from the perspective of joint calibration. 

\subsubsection{About the shape of the kernels}

The present subsection sheds light on the particular role of kernels $K_1$ and $K_2$ in determining the quality of the fit to the market SPX quotes. \\

\begin{figure}[!ht]
    \centering
    \begin{subfigure}[b]{0.49\textwidth}
        \centering
        \includegraphics[width=\textwidth]{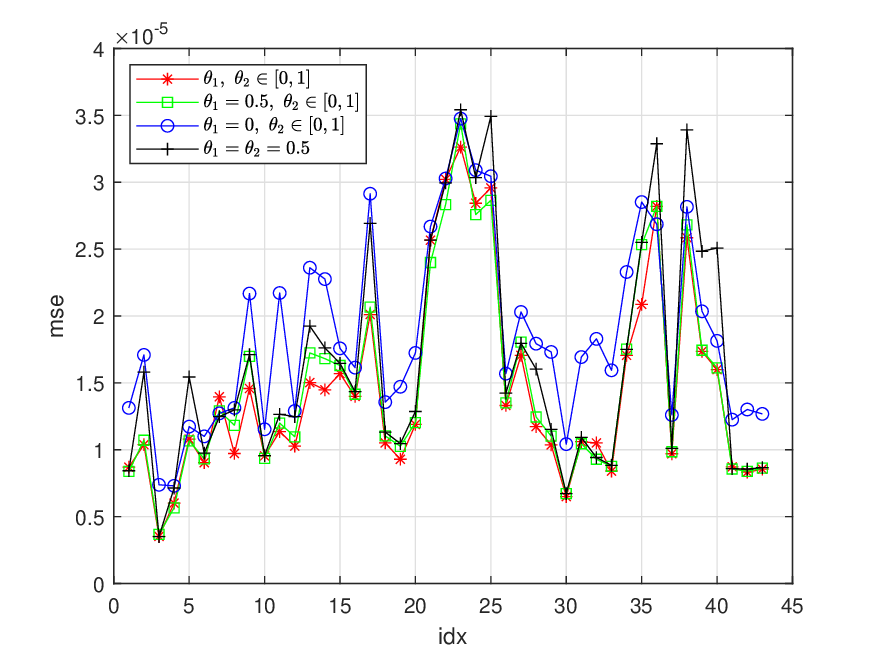}
    \end{subfigure}
    \begin{subfigure}[b]{0.49\textwidth}
        \centering
        \includegraphics[width=\textwidth]{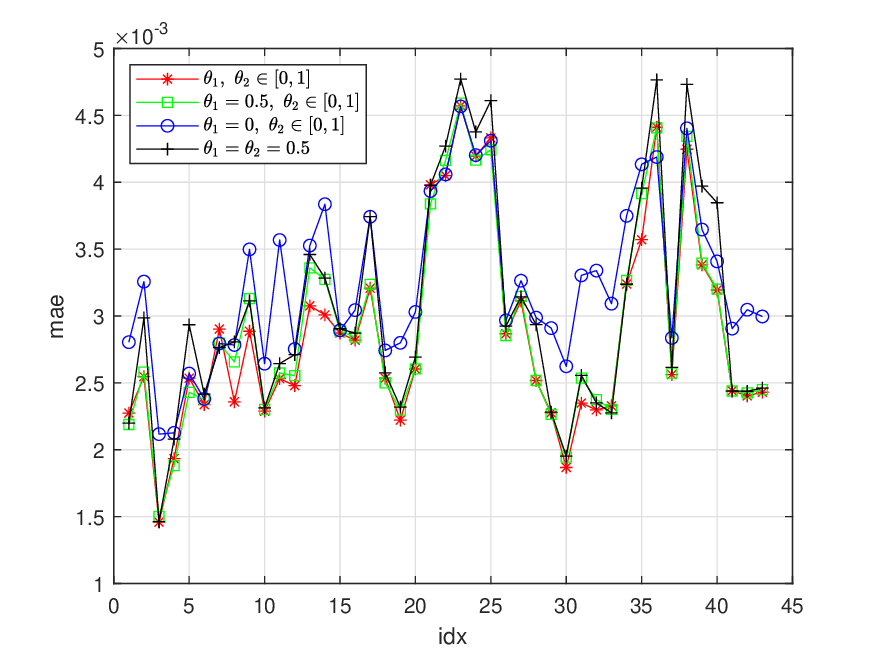}
    \end{subfigure}
    \caption{MSE (left) and MAE (right) over 43 SPX volatility surfaces between 2010 and 2023. Comparing different specifications of the kernels.}
    \label{fig:spx_calib_err_kernels_ts}
\end{figure}

Figure \ref{fig:spx_calib_err_kernels_ts} provides strong evidence for the need for two exponential kernels in both the definition of the trend feature $R_1$ and the activity feature $R_2$. However, mixing by $\theta_1 \in [0,1]$ may not be necessary on $K_1$, as demonstrated by the green points being extremely close to the red ones across the entire sample of dates (except for a couple of points). Fixing $\theta_1 = \theta_2 = 0.5$ is also safe on many occasions, but the evidence shows that restricting $K_2$ by forcing the mixing parameter might be limiting at times (look at the black points) -- contrary to $K_1$.   
It is natural to ask ourselves whether we can make kernel $K_1$ easier, e.g., by reducing it to a single exponential. The answer from the blue points is negative.

\subsection{The joint problem}

We now turn to the joint SPX/VIX calibration problem. For this, we focus on two SPX smiles and one VIX slice (the shortest ones). As noted above, the model is relatively simple, and we cannot expect it to handle the entire surface. Still, in both markets, short-dated options are by far the most liquid. \\

Again, we calibrate using our neural approximations and validate the fits by comparing them with nested Monte Carlo. The VIX benchmark comes with $N^\text{sim}_\text{out}=2^{18}$ outer paths and $N^\text{sim}_\text{inn}=2^{13}$ inner paths. The compliance of the neural fit with the Monte Carlo serves two purposes: 1) it demonstrates that the NN has learned LSMC pricing, and 2) it shows that LSMC is a good approximation for the much more expensive nested simulation.

\subsubsection{October 21, 2009}

\begin{table}[!ht]
    \centering
    \begin{tabular}{|c|c|c|c|c|c|c|c|c|c|}
        \hline
        $\beta_0$ & $\beta_1$ & $\beta_2$ & $\beta_{1,2}$ & $\lambda_{1,0}$ & $\lambda_{1,1}$ & $\theta_1$ & $\lambda_{2,0}$ & $\lambda_{2,1}$ & $\theta_2$ \\ \hline
        0.0840 & -0.2568 & 0.7415 & 0.2078 & 35.57 & 6.99 & 0.8142 & 10.15 & 0.21 & 0.9691 \\ \hline
    \end{tabular}
    \caption[Joint SPX-VIX calibration of the 4-factor Markov PDV model as of October 21, 2009. Parameters.]{Joint SPX-VIX calibration of the 4-factor Markov PDV model as of October 21, 2009. Model parameters as above. Initial values of the factors: $R_{1,0,0}=0.2261$, $R_{1,1,0}=0.4361$, $R_{2,0,0}=0.0281$, $R_{2,1,0}=0.0460$. Futures price from the neural network $F_{nn}=0.2456$. Futures price from nested MC $F_{mc}=0.2461$.}
    \label{tab:joint_params_20091021}
\end{table}

\begin{figure}[!ht]
    \centering
    \begin{subfigure}[b]{0.49\textwidth}
        \centering
        \includegraphics[width=\textwidth]{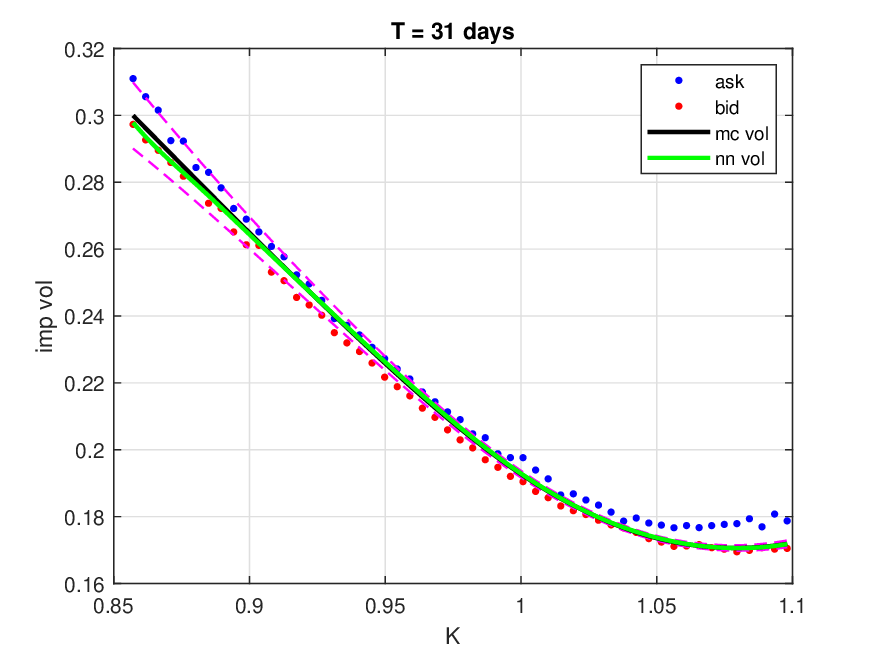}
    \end{subfigure}
    \begin{subfigure}[b]{0.49\textwidth}
        \centering
        \includegraphics[width=\textwidth]{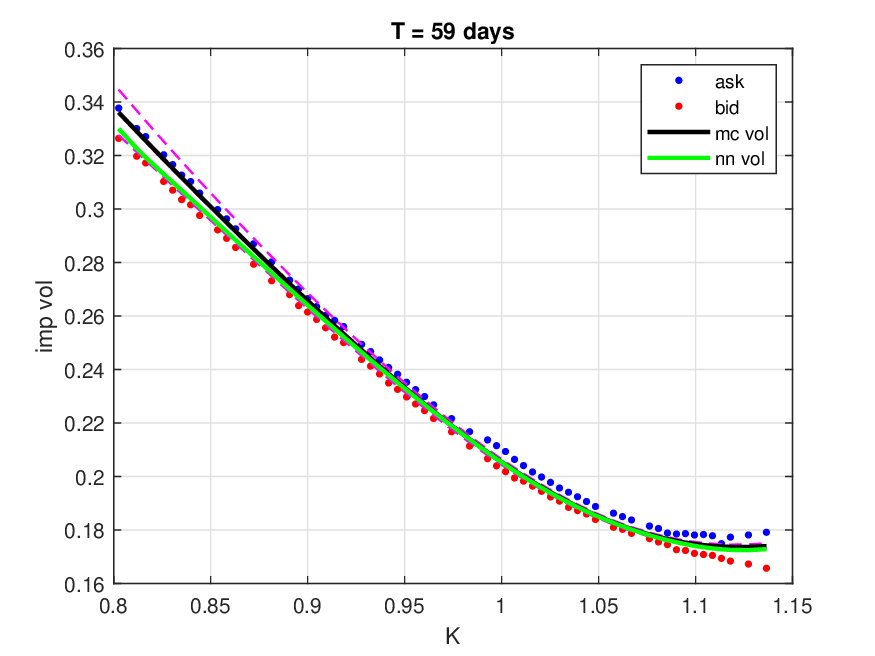}
    \end{subfigure}
    \begin{subfigure}[b]{0.49\textwidth}
        \centering
        \includegraphics[width=\textwidth]{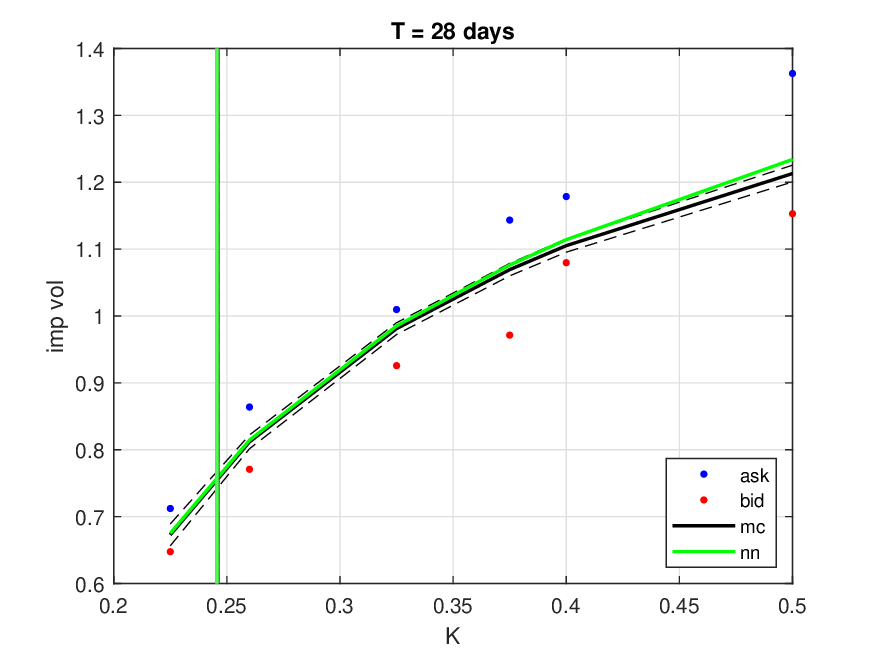}
    \end{subfigure}
    \caption[Joint SPX-VIX calibration of the 4-factor Markov PDV model as of October 21, 2009. Fits.]{Joint SPX-VIX calibration of the 4-factor Markov PDV model as of October 21, 2009. The neural network fit in green. Monte Carlo benchmarks (together with 95\% confidence bands) in (dashed) black.}
    \label{fig:joint_fits_20091021}
\end{figure}

\subsubsection{April 28, 2010}

\begin{table}[!ht]
    \centering
    \begin{tabular}{|c|c|c|c|c|c|c|c|c|c|}
        \hline
        $\beta_0$ & $\beta_1$ & $\beta_2$ & $\beta_{1,2}$ & $\lambda_{1,0}$ & $\lambda_{1,1}$ & $\theta_1$ & $\lambda_{2,0}$ & $\lambda_{2,1}$ & $\theta_2$ \\ \hline
        6.86$\e$-03   & -0.1343   & 0.8774   & 0.1267   & 64.99   & 0.50   & 0.4379   & 36.17   & 3.09   & 0.5435   \\ \hline
    \end{tabular}
    \caption[Joint SPX-VIX calibration of the 4-factor Markov PDV model as of April 28, 2010. Parameters.]{Joint SPX-VIX calibration of the 4-factor Markov PDV model as of April 28, 2010. Model parameters as above. Initial values of the factors: $R_{1,0,0}=-0.5517$, $R_{1,1,0}=0.0525$, $R_{2,0,0}=0.0270$, $R_{2,1,0}=0.0301$. Futures price from the neural network $F_{nn}=0.2084$. Futures price from nested MC $F_{mc}=0.2082$.}
    \label{tab:joint_params_20100428}
\end{table}

\begin{figure}[!ht]
    \centering
    \begin{subfigure}[b]{0.49\textwidth}
        \centering
        \includegraphics[width=\textwidth]{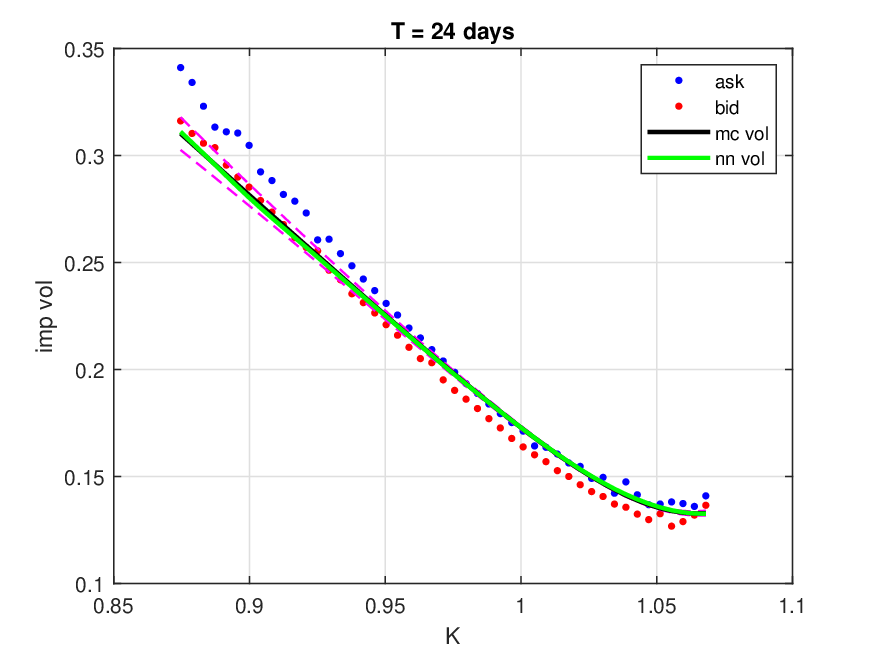}
    \end{subfigure}
    \begin{subfigure}[b]{0.49\textwidth}
        \centering
        \includegraphics[width=\textwidth]{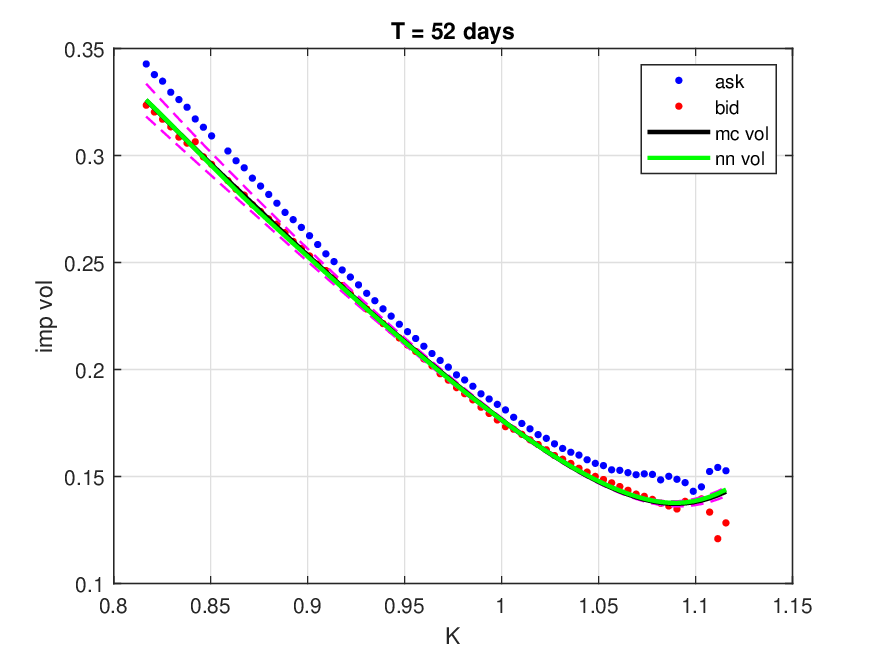}
    \end{subfigure}
    \begin{subfigure}[b]{0.49\textwidth}
        \centering
        \includegraphics[width=\textwidth]{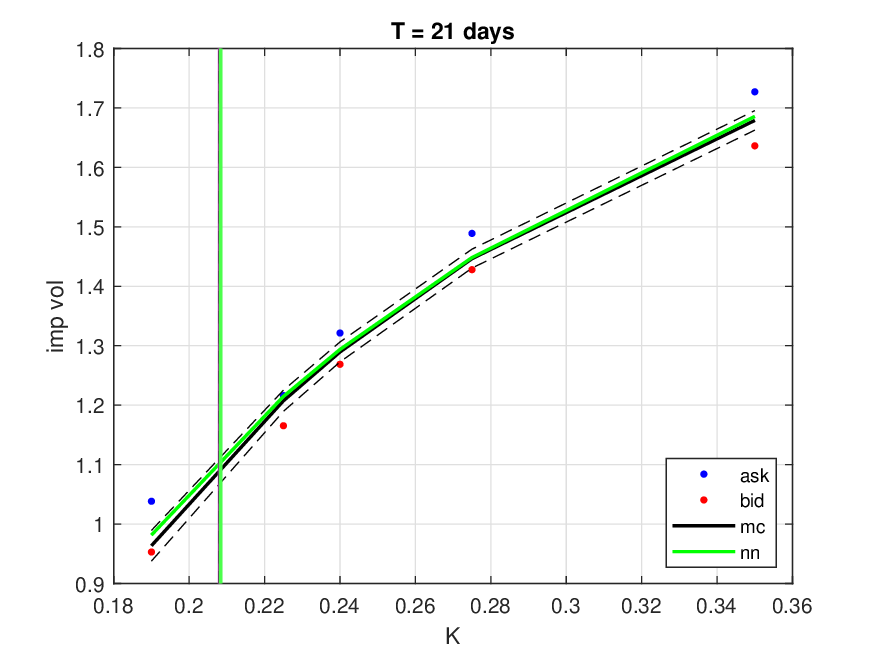}
    \end{subfigure}
    \caption[Joint SPX-VIX calibration of the 4-factor Markov PDV model as of April 28, 2010. Fits.]{Joint SPX-VIX calibration of the 4-factor Markov PDV model as of April 28, 2010. The neural network fit in green. Monte Carlo benchmarks (together with 95\% confidence bands) in (dashed) black.}
    \label{fig:joint_fits_20100428}
\end{figure}

We have looked at two surfaces here. Fits in Figures \ref{fig:joint_fits_20091021} and \ref{fig:joint_fits_20100428}. Model parameters in Tables \ref{tab:joint_params_20091021} and \ref{tab:joint_params_20100428}. \\

The message should be clear at this late stage in the paper. Not only can we optimize model parameters across the entire SPX volatility surface using neural networks, but we can also extend this approach to the joint problem. Pricing of VIX derivatives becomes a matrix-vector product that one can perform in real-time. The cost of the minimization process depends on the number of evaluations of the neural approximation; however, as long as a single call is inexpensive, we are confident that calibration will conclude in a few seconds. Table \ref{tab:times} compares average calibration times in our fully neural approach with the state-of-the-art hybrid method (half NN - half MC) by \cite{GG:25}.

\begin{table}[h!] 
    \centering
    \begin{tabular}{|l|c|c|}
    \hline
    & \textbf{Half NN -- Half MC} & \textbf{Fully Neural} \\
    \hline
    \texttt{calibration time (avg)} & 12 m & 5 s \\
    \texttt{device}           & GPU    & CPU \\
    \texttt{\# cores}         &        & 1 \\
    \hline
    \end{tabular}
    \caption{Average calibration times under our approach (right) as opposed to the state of the art (left).}
    \label{tab:times}
\end{table}    

In conclusion, eliminating the outer simulation is a significant boost to pricing and calibration. It is worth noting that the times we quote refer to a serial procedure. We are not parallelizing over CPU cores at the optimization stage.

\section{Conclusions} \label{sec:concl}

The paper describes a fully fledged neural network that approximates SPX option and VIX derivative prices within the 4-factor Markov PDV model. Two markets require two networks. Both of them take model parameters ($\bm{\theta}$) and contract specifications ($T$, $K$) as input. Then, one can evaluate derivative prices for any strike-maturity pair within the domain of training. Calibration can be performed directly on the market points without any interpolation or extrapolation. This pointwise approach marks an important difference with respect to grid-based methods, as noted in \cite{GJR:20}. We provide users with all the necessary tools to ensure a fast valuation of SPX options and VIX derivatives without restrictions to any calibration grid. We achieve 
the desired speed by properly training the network offline, where we replace computationally cumbersome nested simulations with simple matrix-vector products on top of a limited number of Monte Carlo trajectories. \\

We provide a neural pricer for VIX options and futures and use it in combination with the neural SPX option pricer developed in \cite{BBR:24}
to attack the joint calibration problem. Our focus on the 4-factor Markov PDV model by \cite{GL:23} bridges a gap in the literature, as the state-of-the-art calibration by \cite{GG:25} still relies on online simulation (for the SPX and VIX outer loops in nested simulation). In our approach, the generation phase runs entirely offline. The online calibration consequently gains considerable speed. It now becomes a matter of seconds, which fact is essential for practitioners. \\

The main body of the paper provides extensive evidence of the numerical accuracy of the neural approximations to the pricing functions. In this respect, we emphasize the crucial aspect of reproducing reliable VIX futures prices when calibrating to VIX-implied volatilities. \\

The success of the whole procedure we put into place essentially builds on the quality of the training set. In this respect, the crucial step has been accelerating the generation of VIX samples by leveraging least squares regression and enhancing parameter stability through ridge penalization.

\bibliographystyle{abbrvnat}

\end{document}